\documentclass{article}
\usepackage[utf8]{inputenc}
\usepackage{subcaption}
\usepackage{multirow}
\usepackage{graphicx}
\usepackage{marvosym}
\usepackage{paralist}
\usepackage{hyperref}
\usepackage{amsmath}
\usepackage{authblk}

\DeclareMathOperator*{\argmin}{arg\,min}

\usepackage{xcolor}  
\usepackage{colortbl}

\usepackage{tabularx}
\newcolumntype{Y}{>{\centering\arraybackslash}X}

\usepackage[style=numeric]{biblatex}
\bibliography{references}

\begin{document}
\title{\bf Trav-SHACL: Efficiently Validating Networks of SHACL Constraints}

\author[1,2]{Mónica Figuera}
\author[1,3]{Philipp D. Rohde}
\author[1,3]{Maria-Esther Vidal}

\affil[1]{L3S Research Center, Leibniz University of Hannover, Germany}
\affil[2]{University of Bonn, Germany}
\affil[3]{TIB Leibniz Information Centre for Science and Technology, Hannover, Germany}
\affil[ ]{\normalsize\texttt{monica.figuera@uni-bonn.de, \{philipp.rohde,maria.vidal\}@tib.eu}}

\date{}

\maketitle

\paragraph{\bf Abstract}
Knowledge graphs have emerged as expressive data structures for Web data. Knowledge graph potential and the demand for ecosystems to facilitate their creation, curation, and understanding, is testified in diverse domains, e.g., biomedicine. The Shapes Constraint Language (SHACL) is the W3C recommendation language for integrity constraints over RDF knowledge graphs. Enabling quality assements of knowledge graphs, SHACL is rapidly gaining attention in real-world scenarios. SHACL models integrity constraints as a network of shapes, where a shape contains the constraints to be fullfiled by the same entities. The validation of a SHACL shape schema can face the issue of tractability during validation. To facilitate full adoption, efficient computational methods are required. We present Trav-SHACL, a SHACL engine capable of planning the traversal and execution of a shape schema in a way that invalid entities are detected early and needless validations are minimized. Trav-SHACL reorders the shapes in a shape schema for efficient validation and rewrites target and constraint queries for the fast detection of invalid entities. Trav-SHACL is empirically evaluated on 27 testbeds executed against knowledge graphs of up to 34M triples. Our experimental results suggest that Trav-SHACL exhibits high performance gradually and reduces validation time by a factor of up to 28.93 compared to the state of the art.


\begin{figure}[!ht]
    \centering
    \subfloat[SHACL Shape Network]{
        \includegraphics[scale=.34]{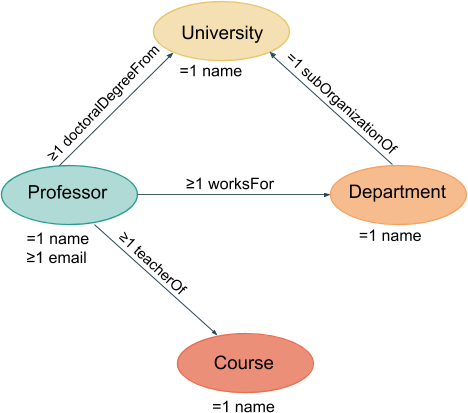}
        \label{fig:motivating-example:network}
    }\hspace*{1cm}
    \subfloat[Random Traversal]{
        \includegraphics[scale=.34]{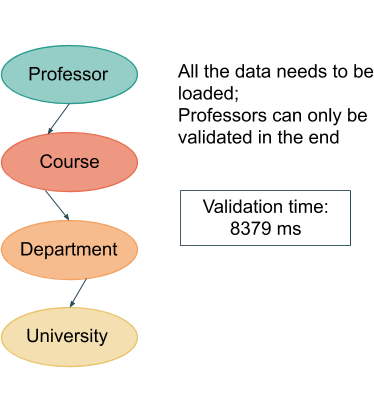}
        \label{fig:motivating-example:random}
    }
    ~\\
    \subfloat[Following Links]{
        \includegraphics[scale=.34]{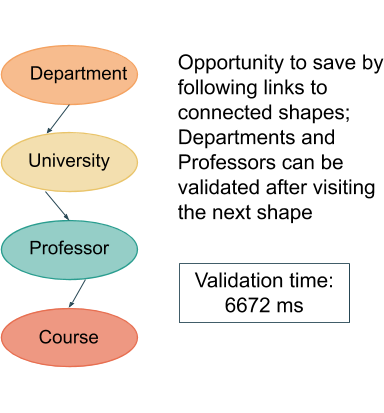}
        \label{fig:motivating-example:links}
    }\hspace*{1cm}
    \subfloat[Exploit Knowledge]{
        \includegraphics[scale=.34]{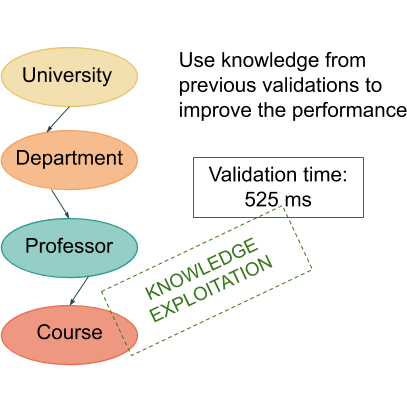}
        \label{fig:motivating-example:knowledge}
    }
    \caption{\textbf{University System Example.} (a) A simple SHACL shape schema for professors, departments, courses, and universities. (b) A random traversal of the network leading to a high validation time. (c) A traversal strategy following links to connected shapes; slight improvement in validation time. (d) A sophisticated traversal strategy that exploits knowledge gained from previous shapes; leading to a reduction in validation time by a factor of 15.96 compared to a random traversal.}
    \label{fig:motivating-example}
\end{figure}

\setcounter{footnote}{0}
\section{Introduction}\label{sec:intro}
Web data keeps enduring an exponential growth rate, and knowledge graphs~\cite{abs-2003-02320} have emerged as expressive data structures that provide a unified view of myriad data sources. Knowledge graphs make possible a holistic description of real-world entities as structured and factual statements. Scientific and industrial communities~\cite{AuerKPKSV18,NoyGJNPT19} are considering them as fundamental building blocks in a new knowledge-driven era of science and technology. The adoption of knowledge graphs in large IT companies~\cite{NoyGJNPT19}, industrial data spaces\footnote{The International Data Space (IDS) \url{https://www.internationaldataspaces.org/}}, and domain-specific applications like in biomedicine~\cite{Nicholson2020} testify not only their potential but also the demand of ecosystems of tools to facilitate their creation, curation, and understanding. In this direction, W3C has actively contributed with standards for declaratively representing knowledge graphs and the whole process of creation and curation. Nevertheless, the full adoption of W3C standards demands the Web community to develop efficient tools to scale up and resist the forecast avalanche of Web data.

The Shapes Constraint Language (SHACL) is the W3C recommendation language for the declarative specification of data quality assessment over RDF knowledge graphs. SHACL is rapidly gaining attention in real-world scenarios, and has been adopted in industrial consortiums (e.g., the International Data Space (IDS)) to represent integrity constraints in reference architectures.
SHACL models integrity constraints as a network of shapes (a.k.a. shape schema).
A shape specifies constraints against the attributes of a specific RDF class, while integrity requirements on properties associating two classes are expressed with links between shapes.
Albeit exhibiting clarity and readability, SHACL shape schemas can face tractability issues during validation. The problem is in general intractable~\cite{Corman2018}, and the current algorithms do not scale well when the size of shape schema or knowledge graph grows. Thus, despite the encouraging evidence about the acceptance of SHACL and other W3C standards, efficient computational tools are required to facilitate full adoption. 

\noindent \textbf{Problem Statement and Objectives.} We address the problem of scaling up the validation of SHACL shape schemas against large knowledge graphs.
Although there are significant contributions (e.g., identification of SHACL tractable fragments~\cite{Corman2018} and algorithms to effectively validate these fragments~\cite{Corman2019}), knowledge graph size imposes strict data management requirements to launch these SHACL engines into real-world scenarios.
We formalize the SHACL validation process as an optimization problem. Given a shape schema, an equivalent one that minimizes the traversal and validation time corresponds to a solution to the problem.  

\noindent \textbf{Our Proposed Solution}. We present Trav-SHACL, a SHACL engine capable of planning a shape schema traversal and executing the shapes in a way that invalid entities are detected early and needless validations are minimized. 
Trav-SHACL implements a two-fold approach. First, it resorts to graph-based metrics -- describing a shape network's connectivity -- to identify a seed shape and the traversal validation strategy. Building on related work, Trav-SHACL implements an algorithm that interleaves data collection from the knowledge graph with constraint validation. These two steps are named \emph{inter-} and \emph{intra-shape} validation, respectively. Trav-SHACL also performs query rewriting techniques that exploit knowledge about the entities (in)validated so far and early identifies new invalidated entities on these entities' neighborhoods.  
We have empirically evaluated Trav-SHACL in 27 testbeds built from knowledge graphs generated using the Lehigh University Benchmark (LUBM)~\cite{Guo2005}.
The study comprises shape schemas and knowledge graphs of various sizes and percentages of invalid entities to ensure the results' reproducibility. Furthermore, we measure the performance of Trav-SHACL in terms of execution time and during an elapsed time period -- or diefficiency. The results indicate savings by a factor of up to 28.93 with respect to SHACL2SPARQL~\cite{Corman2019,Corman2019b}, the state-of-the-art SHACL engine.
Moreover, the query rewriting techniques and interleaved execution of Trav-SHACL allow for an effective forecast of invalid entities and produce the first result ahead of other engines.
More importantly, Trav-SHACL exhibits high-performance continuous behavior and keeps generating results incrementally. 

\noindent \textbf{Contributions}
\begin{inparaenum}[\bf i\upshape)]
    \item Trav-SHACL, a SHACL engine that resorts to query rewriting and interleaved execution strategies to early identify invalid entities and avoid unnecessary constraint validations. 
    \item A set of testbeds that include various parameters that impact the SHACL shape schema validation efficiency.
    \item Empirical evaluation of the performance of nine configurations of Trav-SHACL and two different implementations of SHACL2SPARQL. We run 11 various SHACL engines over 164 constraints validated over nine knowledge graphs whose sizes range from 1M to 34M. Results indicate reductions of execution time by a factor of up to 28.93, while the continuous behavior is considerably improved. Trav-SHACL is available as open-source, and it will be publicly published together with the experimental configuration to ensure reproducibility.
\end{inparaenum}

The remainder of this paper is organized as follows.
\autoref{sec:motivation} motivates our work with an example.
Then \autoref{sec:approach} discusses the planning and execution techniques implemented in Trav-SHACL.
\autoref{sec:eval} empirically evaluates the approach.
Afterwards, \autoref{sec:relatedwork} places our work within the related work.
Finally, we close the paper in \autoref{sec:conclusion} with conclusions and an overview of future work.

\section{Motivating Example}\label{sec:motivation}

Consider a set of SHACL shapes representing constraints over a knowledge graph of a university system, as presented in \autoref{fig:motivating-example:network}.
In SHACL terms, a shape is a set of integrity constraints that apply to the same entities, e.g., the shape called \emph{Professor} in the example.
\emph{Professor}s have precisely one name, at least one e-mail address, at least one doctoral degree from a \emph{University}, i.e., an instance of that shape that meets all constraints, and they work for at least one \emph{Department}.
We call constraints that refer to other shapes \emph{inter-shape constraints}.
Analogous we call constraints that do not refer to other shapes \emph{intra-shape constraints}.
The constraints from the example can be represented with SHACL's \emph{min} and \emph{max constraints} to restrict the minimal and maximal occurrence of a particular pattern, respectively. The constraint on the names of professors is represented as a min and a max constraint, both with the value $1$. The existential constraints can be transformed into min constraints with a minimum of 1.
This example comprises nine min constraints and five max constraints. The nine min constraints are distributed as follows, one for the shape \emph{University}, five for the \emph{Professors}, two for \emph{Departments}, and one for the shape representing \emph{Courses}, respectively. \emph{Universities} are restricted by one max constraint, \emph{Professors} by one, \emph{Departments} by two, and \emph{Courses} by one, respectively.
Shapes can have a target definition, i.e., the shape applies to all entities of a particular class in the RDF data set that is to be validated.
A shape is valid over the data set if and only if all entities in the data set that satisfy the target definition of the shape also fulfill all constraints of the shape.
The strategy followed to traverse the SHACL shapes impacts on the validation time.
To illustrate this issue, consider the strategy presented in \autoref{fig:motivating-example:random}.
In a random traversal order, all the data needs to be loaded.
In the presence of inter-shape constraints, a particular shape, e.g., \emph{Professor}, can only be validated after visiting the shape that is referred to.
Based on the traversal order, this might be a previous shape, the next shape, or even a shape that is scheduled after several others.
To reduce the time necessary to wait until all the information needed is available, the strategy depicted in \autoref{fig:motivating-example:links} follows the links to connected shapes.
This assessment strategy allows for a minor improvement in validation time since some needed information might already be available at a later stage.
In this example, the validation of the shape \emph{Professor} can make use of the validation of the universities and departments.
Following the idea of reusing existing knowledge, if the algorithm keeps track of the valid entities and the violations, this knowledge can be used in the next steps to speed up the validation by identifying invalid entities fast.
The traversal strategy in \autoref{fig:motivating-example:knowledge} follows this approach by starting to validate shapes that are not dependent on other shapes, but other shapes depend on, i.e., the universities in the example.
Nevertheless, the number of constraints to be checked can be dramatically decreased whenever knowledge concerning the validated shapes is exploited to invalidate entities in cascade.
Hence, the quality assessment time of the set of SHACL shapes can be reduced significantly.
As shown in the example, the validation time is sped up by a factor of 15.9 compared to the random traversal.

\section{Trav-SHACL}\label{sec:approach}
\begin{table}[h!]
\caption{\bf Summary of Trav-SHACL Notation}\label{tab:notations}
\normalsize
\centering
{\scriptsize
\begin{tabular}{|m{2.5cm}|m{8.5cm}|}
\hline
\rowcolor{orange!40}
\textbf{Notation} & \textbf{Explanation} \\ \hline
 $\mathcal{S}=\langle S,\textsc{targ},\textsc{def}\rangle$ Shape schema &  $S$ is a set of shape names, $\textsc{targ}$ assigns a SPARQL query to a shape $s$ in $S$, and $\textsc{def}$ maps $s$ to a constraint $\phi$.\\ \hline 
$\mathcal{G}=\langle V_\mathcal{G},E_\mathcal{G} \rangle$  & RDF graph modeling subject, predicate, and object statements. $V_\mathcal{G}$ is a set of subjects and objects, and labelled edges in $E_\mathcal{G}$ represent the RDF triples of $\mathcal{G}$. \\ \hline
$\gamma(\textsc{def}(s))$ & SPARQL query representing the evaluation of $\textsc{def}(s)$.\\ \hline
$[[Q]]^\mathcal{G}$ & Set of mappings from variables in the SPARQL query $Q$ to entities in $\mathcal{G}$ representing the evaluation of $Q$ over $\mathcal{G}$.\\ \hline
$\Phi_\mathcal{S}$ & Dependency graph of the shapes in $\mathcal{S}$.\\ \hline
$\sigma(v,s)$ & Boolean assignment from $V_\mathcal{G}$ and $S$. $\sigma(v,s)=\text{T}$ represents that an entity $v$ belongs to the target of $s$, i.e., there is a variable mapping that includes $v$ in $[[\textsc{targ}(s)]]^\mathcal{G}$; F, otherwise.\\ \hline
$[\phi]^{\mathcal{G},v,\sigma}$ & Boolean function denoting if a constraint $\phi$ is satisfied by an entity $v$ from $\mathcal{G}$ according to an assignment $\sigma$. \\ \hline
$[\mathcal{S}]^\mathcal{G}$ & Entities in $V_\mathcal{G}$ such that exists an assignment $\sigma$, and for each $v$ in $[\mathcal{S}]^\mathcal{G}$, there is a shape $s$ in $S$ and $[\textsc{def}(s)]^{\mathcal{G},v,\sigma}$ is true.\\ \hline
$\text{Time}(\mathcal{S}, \mathcal{G})$ & Cost function representing the time required to evaluate $\mathcal{S}$ over $\mathcal{G}$.\\ \hline
$\sigma_{\textsc{minFix}}^{\mathcal{S},\mathcal{G}}(v,s)$ & Boolean assignment from $V_\mathcal{G}$ and $S$, s.t., $\sigma_{\textsc{minFix}}^{\mathcal{S},\mathcal{G}}(v,s)$ $\equiv$\vspace*{.2em} $[\textsc{def}(s)]^{\mathcal{G},v,\sigma_{\textsc{minFix}}^{\mathcal{S},\mathcal{G}}}$.\\ \hline
$\Gamma_{\mathcal{S},\mathcal{G}}$ & Union of the entities from $\mathcal{G}$ in the result of the evaluation of the SPARQL queries of $\textsc{targ}(.)$ and $\gamma(\textsc{def}(.))$ in $\mathcal{S}$.\\ \hline
\end{tabular}%
}
\end{table}

Trav-SHACL is a data quality assessment engine that resorts to constraints expressed as a shape schema $\mathcal{S}=\langle S,\textsc{targ},\textsc{def}\rangle$ to validate the quality of an RDF graph $\mathcal{G}=\langle V_\mathcal{G},E_\mathcal{G} \rangle$. The evaluation of $\mathcal{S}$ results into a set of entities in $V_\mathcal{G}$ (a.k.a. $[\mathcal{S}]^\mathcal{G}$) that satisfy the constraints in $\mathcal{S}$. Trav-SHACL converts a shape schema $\mathcal{S}$ into an equivalent $\mathcal{S'}=\langle S,\textsc{targ'},\textsc{def}\rangle$ whose validation identifies the same entities but in less time (a.k.a. $\text{Time}(\mathcal{S},\mathcal{G}$)). We formally define this optimization problem and present a heuristic-based approach that identifies low-cost strategies for solving this validation problem. \autoref{tab:notations} summarizes the notation utilized in this section.

\begin{figure}[t]
    \centering
    \includegraphics[width=\linewidth]{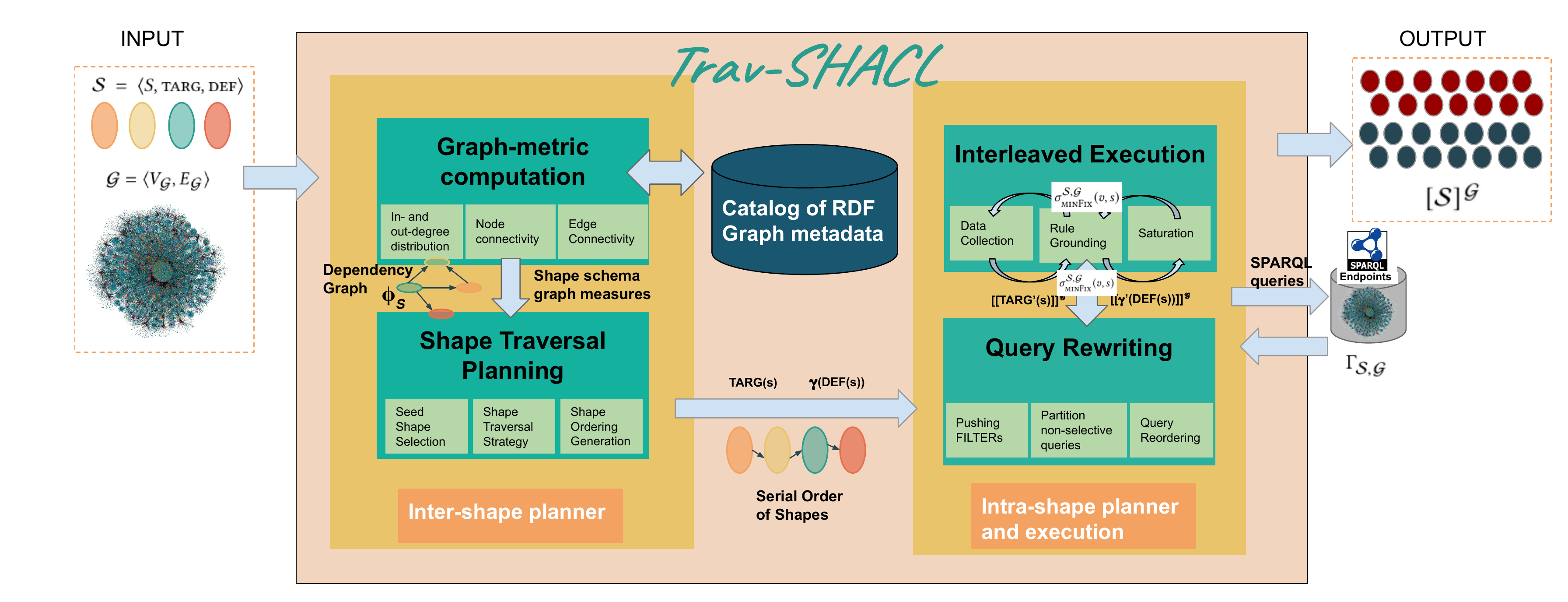}
    \caption{\textbf{The Trav-SHACL Architecture.} Trav-SHACL receives a shape schema $\mathcal{S} = \langle S, \textsc{targ}, \textsc{def} \rangle$ and an RDF graph $\mathcal{G}=\langle V_\mathcal{G},E_\mathcal{G} \rangle$, and outputs $[\mathcal{S}]^\mathcal{G}$, the entities in $V_\mathcal{G}$ that satisfy shapes in $\mathcal{S}$. The inter-shape planner resorts to graph metrics computed on the dependency graph $\Phi_\mathcal{S}$; it orders the $\mathcal{S}$ shapes in a way that invalid entities are identified sooner. The intra-shape planner and execution optimizes the queries $\textsc{targ}$ and $\gamma(\textsc{def}(s))$ at the time $\mathcal{S}$ is traversed. So-far (in)validated entities are considered to filter out entities linked to these entities; query rewriting decisions (e.g., pushing filters, partitioning of non-selective queries, and query reordering) are made based on invalid entities' cardinalities and query selectivity. Rewritten queries are executed against SPARQL endpoints. Query answers $[[\textsc{targ}(s)]]^\mathcal{G}$ and $[[\gamma(\textsc{def}(s))]]^\mathcal{G}$, and truth value assignments $\sigma_{\textsc{minFix}}^{\mathcal{S},\mathcal{G}}$, are exchanged during query rewriting and interleaved execution. They are utilized -- in a bottom-up fashion -- for constraint rule grounding and saturation. The intra-shape planner and execution component runs until a fixed-point in $\sigma_{\textsc{minFix}}^{\mathcal{S},\mathcal{G}}$ is reached.} 
    \label{fig:architecture}
\end{figure}

\subsection{Preliminaries}
The Resource Description Framework (RDF) is the W3C standard for publishing and exchanging data over the web. It is commonly utilized to represent knowledge graphs as a set of triples that consist of three parts: \begin{inparaenum}[\bfseries (i)]
     \item subject - an entity or resource,
     \item predicate - a relation between subject and object, and
     \item object - an entity or resource\end{inparaenum}, just like a sentence.
Subjects and predicates are always \emph{Universal Resource Identifiers} (URI).
In addition, objects can also be represented as a literal instead of an URI to use data formats like strings, integers, or dates. In an RDF graph $\mathcal{G}=\langle V_\mathcal{G},E_\mathcal{G} \rangle$, nodes correspond to subjects and objects, while predicates are the label of directed edges from subjects to objects.
The Shapes Constraint Language (SHACL) is the W3C recommendation language for representing integrity constraints over an RDF graph; we follow the abstract syntax and semantics defined by Corman et al.~\cite{Corman2018,Corman2019}.
A \emph{shape schema} defined as $\mathcal{S}=\langle S,\textsc{targ},\textsc{def}\rangle$, represents the set $S$ of \emph{shape names}, and two functions $\textsc{targ}$ and $\textsc{def}$ that map a shape with a target query and with a constraint, respectively.
A target query states the 
RDF class -- in an RDF graph $\mathcal{G}$ -- of the entities for which the corresponding shape will be validated. We assume target queries are expressed in SPARQL. 
The result of the evaluation of a target query $Q$ over $\mathcal{G}$ (a.k.a. $[[Q]]^\mathcal{G}$) corresponds to a set of mappings from the variables in $Q$ to the entities in $\mathcal{G}$ that satisfy the graph patterns in $Q$ \cite{PerezAG09}. In case $Q$ only includes the variable $?x$ in the select clause, $[[Q]]^\mathcal{G}$=$\{\{(?x,v_1)\},\dots,(?x,v_m)\}$, where $v_1,\dots,v_m$ are entities in $V_\mathcal{G}$. A dependency graph $\Phi_\mathcal{S}$ of a shape schema $\mathcal{S}$ is a directed graph where shapes in $\mathcal{S}$ are represented as nodes, and an edge $(s_i,s_j)$ indicates that $s_j$ appears in the constraint of $s_i$.
An assignment $\sigma$ is a function that assigns a Boolean value to the entities $v$ in $V_\mathcal{G}$ and the shapes in $\mathcal{S}$.
The interpretation of a constraint $\phi$ of a shape $s$, in an entity $v$ in $V_\mathcal{G}$ according to an assignment $\sigma$ (a.k.a. $[\phi]^{\mathcal{G},v,\sigma}$), is a Boolean function that indicates if $v$ satisfies $\phi$ given $\sigma$; $[\phi]^{\mathcal{G},v,\sigma}$ is inductively defined on the structure of $\phi$, where the base case corresponds to the value of $\sigma(v,s)$.
The entities in $V_\mathcal{G}$ satisfy a shape schema $\mathcal{S}$ (a.k.a. $[\mathcal{S}]^\mathcal{G}$) iff there is an assignment $\sigma$ and a shape $s$ in $\mathcal{S}$, and $[\textsc{def}(s)]^{\mathcal{G},v,\sigma}$ is true. In general the problem determines if the entities of an RDF graph $\mathcal{G}$ satisfy a shape schema $\mathcal{S}$ in a given assignment $\sigma$ in NP-complete \cite{Corman2018}. Nevertheless, Corman et al. \cite{Corman2019} have identified three fragments of SHACL that are tractable; $\mathcal{L}^{\text{non-rec}}$ only enables non-recursive shapes, $\mathcal{L}^{\text{s}}$ does not allow negations through recursive shapes, and $\mathcal{L}^{+}_{\lor}$ does not allow negations but disjunction. More importantly, Corman et al. \cite{Corman2019} propose a computational method that performs the inference process required to construct the set of entities in an RDF graph $\mathcal{G}$ that satisfies a shape schema $\mathcal{S}$. This computational method is ground on the results of deductive databases \cite{CeriGT89} to compute the minimal model of the constraints of the shapes in $\mathcal{S}$ for the entities in $\mathcal{G}$ that correspond to the instantiations of the target queries of these shapes. This minimal model is defined in terms of the fixed-point assignment $\sigma_{\textsc{minFix}}^{\mathcal{S},\mathcal{G}}$, that assigns the same truth value to an entity $v$ in a shape $s$ than the value of satisfaction of $v$ in the constraint of $s$ according to $\sigma_{\textsc{minFix}}^{\mathcal{S},\mathcal{G}}$, i.e.,  $\sigma_{\textsc{minFix}}^{\mathcal{S},\mathcal{G}}(v,s)$ $\equiv$ $[\textsc{def}(s)]^{\mathcal{G},v,\sigma_{\textsc{minFix}}^{\mathcal{S},\mathcal{G}}}$. The minimal model for the fragments $\mathcal{L}^{\text{non-rec}}$, $\mathcal{L}^{\text{s}}$, and $\mathcal{L}^{+}_{\lor}$, $\sigma_{\textsc{minFix}}^{\mathcal{S},\mathcal{G}}$ can be computed in polynomial time in the size of the result of all the queries mapped by $\textsc{targ}(.)$ and that defined the constraints assigned by $\textsc{def}(.)$; the set with the union of all these entities is named $\Gamma_{\mathcal{S},\mathcal{G}}$. 
We propose query optimization techniques that exploit knowledge about the invalid entities identified during the execution of the shapes evaluated so far, as well as the semantics encoded in the RDF graph to rewrite the queries in $\textsc{targ}(.)$ and $\textsc{def}(.)$. Thus, the set of the entities retrieved from the RDF graph that will invalidate the shape schema is minimized. 

\subsection{Problem Statement and Proposed Solution}
\noindent \textbf{Problem Statement:}
Given an RDF graph $\mathcal{G}=\langle V_\mathcal{G},E_\mathcal{G} \rangle$  and a shape schema $\mathcal{S} = \langle S, \textsc{targ}, \textsc{def} \rangle$, the problem of SHACL executing over $\mathcal{G}$ is to find a shape schema $\mathcal{S'} = \langle S, \textsc{targ'}, \textsc{def} \rangle$ that meets the following conditions:
\begin{itemize}
\item $\mathcal{S}$ and $\mathcal{S'}$ are equivalent when evaluated over $\mathcal{G}$. The set of entities in $\mathcal{G}$ that validate 
$\mathcal{S}$ and $\mathcal{S'}$ is the same, i.e., 
$[\mathcal{S}]^\mathcal{G}$=$[\mathcal{S'}]^\mathcal{G}$.
\item The time required to evaluate $\mathcal{S'}$ is minimal, i.e., if $\mathcal{Z}_{\mathcal{G},\mathcal{S}}$ is the set of all the shape schema equivalent to $\mathcal{S}$, then $\mathcal{S'}$ is the schema in  $\mathcal{Z}_{\mathcal{G},\mathcal{S}}$ that minimizes the evaluation time.

\begin{equation} \label{eq:3}
\mathcal{S'}=\argmin\limits_{\mathcal{S'} \in \mathcal{Z}_{\mathcal{G},\mathcal{S}}} \text{Time}(\mathcal{S'}, \mathcal{G})
\end{equation}
\end{itemize}

\noindent \textbf{Solution.} 
Trav-SHACL implements a heuristic-based approach. It relies on the assumption that the minimal retrieval of the entities required to validate a shape $s$ leads to collecting only the entities needed to assess the shapes mentioned in the constraint of $s$, i.e., $\textsc{def}(s)$.
Trav-SHACL follows a two-fold strategy that is guided by heuristics to perform inter- and intra-shape optimizations.
The shapes of $\mathcal{S}$ are reordered; the inter-shape optimizations aim at deciding a traversal strategy where the shapes that invalidate the highest number of entities are validated first. Then, $\mathcal{S}$ is executed following the selected order, and intra-shape optimizations are performed on the fly. According to the number of invalid entities identified during the execution of the previous executed shapes, Trav-SHACL rewrites the target and constraint queries, i.e., $\textsc{targ}(s)$ and $\gamma(\textsc{def}(s))$, 
to filter out entities linked to the entities invalidated so far. As soon as a query answer is retrieved, the collected entities are used to ground the rules that represent $\textsc{def}(s)$. Grounded rules fire a bottom-up evaluation process, named \emph{saturation}, to generate new truth values of entities in $\sigma_{\textsc{minFix}}^{\mathcal{S},\mathcal{G}}$. Saturation is interleaved with data collection and grounding; it finalizes when a fixed-point on $\sigma_{\textsc{minFix}}^{\mathcal{S},\mathcal{G}}$ is reached. Shape reordering together with the execution of the rewritten queries conduce to grounding a small number of constraint rules. More importantly, the interleaved evaluation of the saturation process enables the inference of invalid entities (i.e., false assignments in $\sigma_{\textsc{minFix}}^{\mathcal{S},\mathcal{G}}$) as soon as these entities are collected. As a result, the execution time of the new shape schema $\mathcal{S'}$, $\text{Time}(\mathcal{S'}, \mathcal{G})$, is minimized, and $[\mathcal{S}]^\mathcal{G}$ is created incrementally.   

\begin{figure}[!ht]
    \centering
    \includegraphics[width=\linewidth]{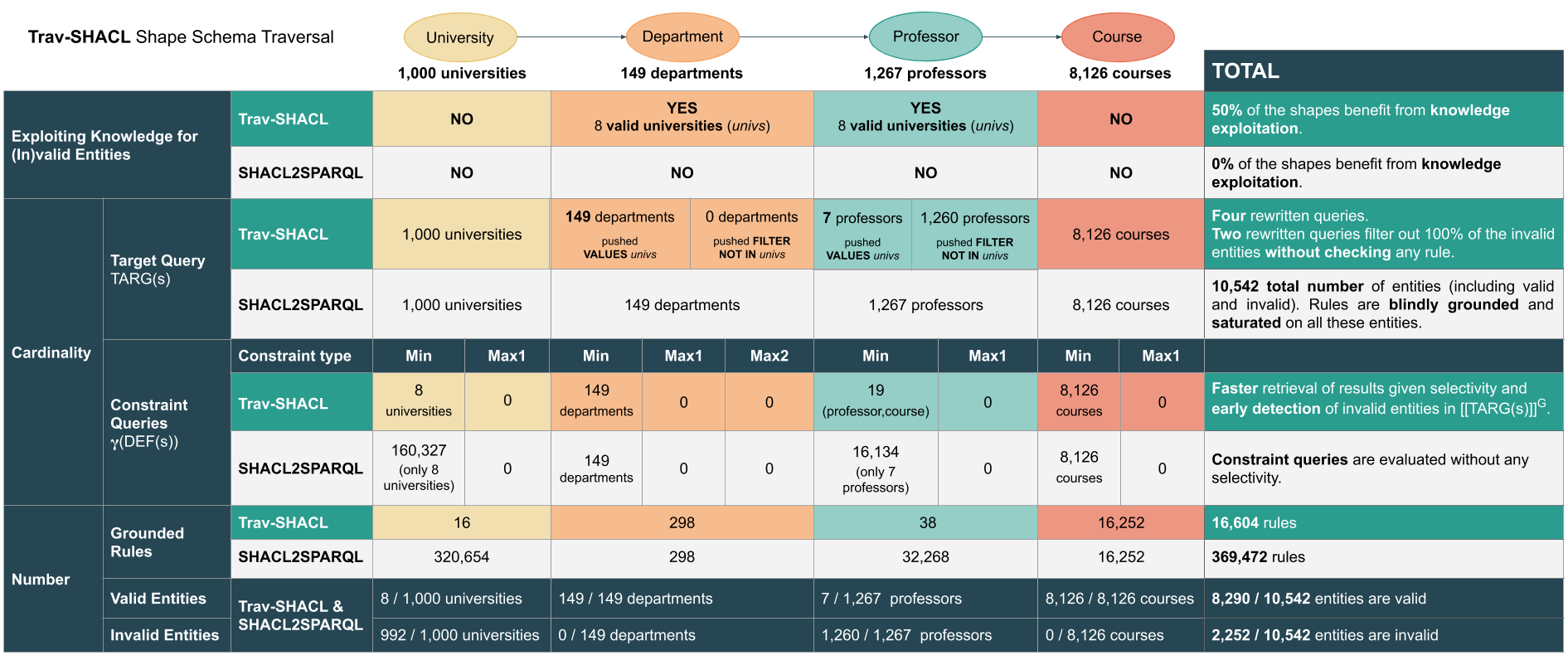}
    \caption[Running example]{\textbf{Running example.} Numbers retrieved for the shape schema presented in \autoref{fig:motivating-example:network}, following Trav-SHACL's traversal order which allows for knowledge exploitation, as depicted in \autoref{fig:motivating-example:knowledge}. Assume SHACL2SPARQL follows the same traversal order. Both engines (in)validate 10,542 entities in total. Nevertheless, Trav-SHACL computes a factor of 22.25 less grounded rules in memory by making use of its \emph{Inter-shape planner} and \emph{Intra-shape planner and execution} components\footnotemark.}
    \label{fig:run_motivating_ex}
\end{figure}

\subsection{The Trav-SHACL Architecture} 
\autoref{fig:architecture} depicts the Trav-SHACL architecture. Given a shape schema $\mathcal{S} = \langle S, \textsc{targ}, \textsc{def} \rangle$ and an RDF graph $\mathcal{G}=\langle V_\mathcal{G},E_\mathcal{G} \rangle$, Trav-SHACL outputs the set of entities in $V_\mathcal{G}$ that satisfy $\mathcal{S}$. Trav-SHACL follows a two-fold approach composed of two main components: \begin{inparaenum}[\bf i\upshape)]
    \item \textit{Inter-shape planner} and 
    \item \textit{Intra-shape planner and execution}. 
\end{inparaenum}
In a first stage, measures -- computed from the dependency graph 
$\Phi_\mathcal{S}$ -- together with statistics about $\mathcal{G}$, feed the heuristic-based approach implemented by the \textit{Inter-shape planner}. In the second stage, the \textit{Intra-shape planner and execution} component rewrites and executes SPARQL queries; it also infers from the collected answers, the entities that validate $\mathcal{S}$. Both components are detailed next. 
~\\
\footnotetext{All min constraints are evaluated in one query, whereas only one max constraint per query is allowed, as formalized by Corman et al. \cite{Corman2019}.}\noindent \textbf{Inter-shape planner}. Trav-SHACL exploits graph-based measures (i.e., in- and out-degree distributions) to determine the connectivity of a shape in the dependency graph $\Phi_\mathcal{S}$ of $\mathcal{S}$. As a result, Trav-SHACL decides the traversal's seed shape and the best search strategy, e.g., depth-first search (DFS) or breadth-first search (BFS). The natural intuition that shapes $s$ with high in-degree values should be evaluated first, enables to considerably reduce the number of retrieved entities during the evaluation of the neighbors of $s$, whenever a large number of entities invalidate $s$.
The seed shape selection is guided by heuristics. First, all shapes $s$ of $S$ that have an empty $\textsc{targ}(s)$, i.e., no class in the RDF graph is assigned to the shape, are discarded from the possible seed shapes. Following the above mentioned intuition, Trav-SHACL selects -- from this list of possible shapes -- the one with the highest in-degree as the seed shape of the traversal.
Assuming that there are still at least two shapes that qualify for the seed shape, the one with the most constraints is chosen based on the intuition that more constraints to be met by a single entity increase the chances of a higher number of invalidated entities. In the example in \autoref{fig:motivating-example}, all shapes have a target definition, so no shape is discarded. The \emph{University} shape has the highest in-degree, i.e., the reference from \emph{Professor} and \emph{Department} to it. Therefore, the other shapes are omitted. Since only one possible seed shape is left, the heuristic considering the number of constraints is not taken into account in this example.
Given a seed shape $s$, the dependency graph $\Phi_\mathcal{S}$, and the traversal strategy (e.g., DFS or BFS), Trav-SHACL's shape ordering generation starts applying the traversal strategy on $\Phi_\mathcal{S}$ at $s$ ignoring edge directions in order to create an enumeration of the shapes in $\mathcal{S}$. This enumeration is a preordering of the vertices in $\Phi_\mathcal{S}$. The approach keeps track of all the nodes in $\Phi_\mathcal{S}$ that are not yet visited and the ones that are visited already. In the case of recursions, when reaching an already visited node $n$, the search continues with the first unvisited neighbor of $n$ in $\Phi_\mathcal{S}$. If there is no such neighbor, the first node in the list of not yet visited nodes is used for continuation. Since in this approach shapes are not visited more than once, the complexity is equal to the worst-case complexity of depth-first search for explicit graphs traversed without repetition, i.e., $\mathcal{O}(|V|+|E|)$ where $V$ are the vertices in $\Phi_\mathcal{S}$ and $E$ the edges in $\Phi_\mathcal{S}$. Trav-SHACL starts to explore the dependency graph $\Phi_\mathcal{S}$ from \emph{University}. The shape is connected to \emph{Professor} and \emph{Department}. Due to the internal representation of the graph, the second node visited is the \emph{Department} node. From \emph{Department} there is an edge to the unvisited node \emph{Professor}, hence, the third shape to be evaluated are the professors. In the neighborhood of \emph{Professor} there is only one unvisited node left, i.e., \emph{Course}. Once \emph{Course} is scheduled to be evaluated at fourth position, all nodes in the dependency graph $\Phi_\mathcal{S}$ have been visited and the enumeration of all nodes $n$ in $\Phi_\mathcal{S}$ is complete. The final traversal order is \emph{University}, \emph{Department}, \emph{Professor}, \emph{Course} and, therefore, the one that \emph{exploits knowledge}  in \autoref{fig:motivating-example:knowledge}.
~\\
\noindent\textbf{Intra-shape planner and execution.} 
Once a traversal is decided, Trav-SHACL starts the shape schema's execution and performs \emph{query rewriting} and \emph{interleaved executions} while the shape schema is traversed.  
The \emph{query rewriting} component performs intra-shape optimizations to increase the selectivity of both target and constraint queries. During \emph{Pushing FILTERs}, the list of entities (in)validated so far by the neighbor shapes are used as filters in the queries $\textsc{targ}(s)$ and $\gamma(\textsc{def}(s))$. Trav-SHACL pushes filters down by making use of the SPARQL VALUES and FILTER NOT IN clauses. Thus, for every shape $s$ in $\mathcal{S}$, and given a list of entities validated per neighboring shape in the dependency graph $\Phi_\mathcal{S}$, Trav-SHACL prioritizes the smallest list -- valid or invalid entities -- to be included in a query filter. 
The \emph{Partition of non-selective queries} is applied whenever the cardinality of a query overpasses the SPARQL endpoint limit of maximal answers. Query offsets are evaluated to continuously retrieve slices of the answers; the SPARQL modifiers LIMIT and OFFSET are defined according to thresholds specified in the configuration of Trav-SHACL. Furthermore, if the list to be included as a query filter is very large, and the rewritten query exceeds the maximum number of characters allowed by a SPARQL endpoint, the query is rewritten into several queries, and the union of the answers of these queries corresponds to the answer of the original one.  The maximum number of rewritten queries is given by a threshold set up according to the configuration of Trav-SHACL. If the number of generated queries is longer than the threshold, the query rewriting is not applied to avoid overhead, and only one query is generated. Lastly, the \emph{Query reordering} aims to execute the most selective queries to increase the validation process' continuous behavior.

Concurrently, the \emph{Interleaved execution} component interleaves the verification of the constraints with the execution of the queries. Thus, entities can be validated as soon as they are retrieved, allowing Trav-SHACL to produce results incrementally.
First, each entity $v$ in $[[\textsc{targ}(s)]]^\mathcal{G}$ is used to ground all the constraints of $\textsc{def}(s)$.
Trav-SHACL follows the approach proposed by~\cite{Corman2018,Corman2019}, and represents the constraints as a theory $\mathcal{T}$ of safe stratified rules. These rules are of the form $l_0 \land \dots l_n \implies s(v)$, where each $l_i$ corresponds to $s_i(v_i)$ or $\neg s_i(v_i)$, $s_i$ is a shape in $\mathcal{S}$, and $v_i$ belongs to $V_\mathcal{G}$. $\mathcal{T}$ is built in the way that every model of $\mathcal{T}$ in the entities of $V_\mathcal{G}$ corresponds to the entities in $[\mathcal{S}]^\mathcal{G}$, i.e., the entities that satisfy $\mathcal{S}$. During $\emph{saturation}$, grounded rules are utilized to infer which entities validate $\mathcal{T}$. Rules representing the constraints in $\textsc{def}(s)$ are validated, and once an entity $v$ in $[[\textsc{targ}(s)]]^\mathcal{G}$ invalidates a constraint in $\gamma(\textsc{def}(s))$, Trav-SHACL skips the evaluation of the remaining rules where $s(v)$ appears in the body. Trav-SHACL also adds a false value to $s(v)$ in $\sigma_{\textsc{minFix}}^{\mathcal{S},\mathcal{G}}$. The component \emph{Intra-shape planner and execution} keeps the execution of the
\emph{Query Rewriting} and \emph{Interleaved execution} components until a fixed-point in $\sigma_{\textsc{minFix}}^{\mathcal{S},\mathcal{G}}$. 
\autoref{fig:motivating-example:knowledge} and its corresponding running example in \autoref{fig:run_motivating_ex} illustrate the result of executing the \textit{Inter-shape planner} and \textit{Intra-shape planner and execution}. Further, \autoref{fig:run_motivating_ex} reports on the results of the state-of-the-art SHACL2SPARQL empowered with the traversal identified by Trav-SHACL.
Trav-SHACL not only traverses the shapes in the selected order, but also pushes into the SPARQL queries the corresponding filters to avoid interleaving entities linked to already invalidated ones. This is depicted in \autoref{fig:run_motivating_ex}, where using the knowledge of the eight valid universities allows to rewrite $\textsc{targ}(\text{Professor})$ to filter out 1,260 invalid entities from grounding. This inter- and intra-shape strategy collects 9,282 entities that result in grounding 16,604 rules versus 369,472 rules grounded by SHACL2SPARQL. Consequently, Trav-SHACL execution time is 525ms, with savings of a factor of 22.25 in the number of grounded rules with respect to SHACL2SPARQL. Going back to the motivating example,  
in the shape schema $\mathcal{S'}$ and $\mathcal{S''}$ in \autoref{fig:motivating-example:random} and \autoref{fig:motivating-example:links} only inter-shape traversal decisions have been taken into account. The number of retrieved entities of $\mathcal{S}$ (i.e., $\Gamma_{\mathcal{S},\mathcal{G}}$) is 10,542 and remains the same in  $\Gamma_{\mathcal{S'},\mathcal{G}}$ and  $\Gamma_{\mathcal{S''},\mathcal{G}}$. Nevertheless, because of the ordering in which the shapes are evaluated, entities are invalidated as soon as they are retrieved, and the constraints of $\mathcal{S'}$ and $\mathcal{S''}$ are grounded and validated in less number of entities. Thus, $\Gamma_{\mathcal{S'},\mathcal{G}}$ and  $\Gamma_{\mathcal{S''},\mathcal{G}}$ ground all the constraints in the retrieved entities and end up validating 369,472 and 337,430 rules, respectively. As a result, differences in $\text{Time}(\mathcal{S'},\mathcal{G})$ and $\text{Time}(\mathcal{S''},\mathcal{G})$ are observed.

\section{Empirical Evaluation}\label{sec:eval}

\begin{table*}[t!]
\caption{\textbf{Data Statistics.} \#triples - number of triples in the KG, \#subjects - number of distinct subjects in the KG, \%inv - percentage of invalid entities in the KG based on validating the network of shapes, C - number of constraints in the network, $\overline{\text{inter}}$ - average checks for each inter-constraint, $\overline{\text{intra}}$ - average checks for each intra-constraint}
\label{tab:datasets}
    \centering
{\tiny
\begin{tabular}{|c|c|c|c|c|c|c|c|c|}
\hline
\rowcolor{orange!50} 
\multicolumn{9}{|c|}{\textbf{Shape Schema 1: $|S|=3$, C=16}}\\
\cline{1-9}
\rowcolor{orange!20}
\multicolumn{3}{|c|}{Small Knowledge Graphs (SKGs)} & \multicolumn{3}{|c|}{Medium Knowledge Graphs (MKGs)} & \multicolumn{3}{|c|}{Large Knowledge Graphs (LKGs)}\\
\textbf{D1} & \textbf{D2} & \textbf{D3} & \textbf{D4} & \textbf{D5} & \textbf{D6} &  \textbf{D7} & \textbf{D8} & \textbf{D9}\\
\multicolumn{1}{|c}{\textbf{\#triples:}} & \multicolumn{2}{c|}{1,001,420} & \multicolumn{1}{|c}{\textbf{\#triples:}} & \multicolumn{2}{c|}{4,257,051} &
\multicolumn{1}{|c}{\textbf{\#triples:}} & \multicolumn{2}{c|}{34,095,887}\\
\multicolumn{1}{|c}{\textbf{\#subjects:}} & \multicolumn{2}{c|}{163,553} & \multicolumn{1}{|c}{\textbf{\#subjects:}} & \multicolumn{2}{c|}{693,270} &
\multicolumn{1}{|c}{\textbf{\#subjects:}} & \multicolumn{2}{c|}{5,543,986}\\
\multicolumn{1}{|c}{$\overline{\textbf{inter}}$} & \multicolumn{2}{c|}{947.57} & \multicolumn{1}{|c}{$\overline{\textbf{inter}}$} & \multicolumn{2}{c|}{4,072.86} &
\multicolumn{1}{|c}{$\overline{\textbf{inter}}$} & \multicolumn{2}{c|}{32,500.71}\\
\multicolumn{1}{|c}{$\overline{\textbf{intra}}$} & \multicolumn{2}{c|}{959.22} & \multicolumn{1}{|c}{$\overline{\textbf{intra}}$} & \multicolumn{2}{c|}{3,390.00} &
\multicolumn{1}{|c}{$\overline{\textbf{intra}}$} & \multicolumn{2}{c|}{25,500.56}\\
\cline{1-6} 
\rowcolor{orange!20}
\multicolumn{3}{|c|}{SKGs Invalid Entities \textbf{\%inv}}& \multicolumn{3}{|c|}{MKGs Invalid Entities \textbf{\%inv}} & \multicolumn{3}{|c|}{LKGs Invalid Entities \textbf{\%inv}}\\
\textbf{D1} & \textbf{D2} & \textbf{D3} & \textbf{D4} & \textbf{D5} & \textbf{D6} &  \textbf{D7} & \textbf{D8} & \textbf{D9}\\
\text{75.83\%} & \text{87.42\%} & \text{92.22\%} & \text{63.90\%} & \text{81.24\%} & \text{88.51\%} & \text{59.37\%} & \text{78.22\%} & \text{86.86\%} \\
\cline{1-9}
\hline
\rowcolor{orange!50} 
\multicolumn{9}{|c|}{\textbf{Shape Schema 2: $|S|=7$, C=36}}\\
\cline{1-9}
\rowcolor{orange!20}
\multicolumn{3}{|c|}{Small Knowledge Graphs (SKGs)} & \multicolumn{3}{|c|}{Medium Knowledge Graphs (MKGs)} & \multicolumn{3}{|c|}{Large Knowledge Graphs (LKGs)}\\
\textbf{D10} & \textbf{D11} & \textbf{D12} & \textbf{D13} & \textbf{D14} & \textbf{D15} &  \textbf{D16} & \textbf{D17} & \textbf{D18}\\
\multicolumn{1}{|c}{\textbf{\#triples:}} & \multicolumn{2}{c|}{1,001,420} & \multicolumn{1}{|c}{\textbf{\#triples:}} & \multicolumn{2}{c|}{4,257,051} &
\multicolumn{1}{|c}{\textbf{\#triples:}} & \multicolumn{2}{c|}{34,095,887}\\
\multicolumn{1}{|c}{\textbf{\#subjects:}} & \multicolumn{2}{c|}{163,553} & \multicolumn{1}{|c}{\textbf{\#subjects:}} & \multicolumn{2}{c|}{693,270} &
\multicolumn{1}{|c}{\textbf{\#subjects:}} & \multicolumn{2}{c|}{5,543,986}\\
\multicolumn{1}{|c}{$\overline{\textbf{inter}}$} & \multicolumn{2}{c|}{16,819.07} & \multicolumn{1}{|c}{$\overline{\textbf{inter}}$} & \multicolumn{2}{c|}{71,414.00} &
\multicolumn{1}{|c}{$\overline{\textbf{inter}}$} & \multicolumn{2}{c|}{573,048.13}\\
\multicolumn{1}{|c}{$\overline{\textbf{intra}}$} & \multicolumn{2}{c|}{27,217.33} & \multicolumn{1}{|c}{$\overline{\textbf{intra}}$} & \multicolumn{2}{c|}{115,383.00} &
\multicolumn{1}{|c}{$\overline{\textbf{intra}}$} & \multicolumn{2}{c|}{924,155.05}\\
\cline{1-6} 
\rowcolor{orange!20}
\multicolumn{3}{|c|}{SKGs Invalid Entities \textbf{\%inv}}& \multicolumn{3}{|c|}{MKGs Invalid Entities \textbf{\%inv}} & \multicolumn{3}{|c|}{LKGs Invalid Entities \textbf{\%inv}}\\
\textbf{D10} & \textbf{D11} & \textbf{D12} & \textbf{D13} & \textbf{D14} & \textbf{D15} &  \textbf{D16} & \textbf{D17} & \textbf{D18}\\
\text{1.38\%} & \text{1.59\%} & \text{1.68\%} & \text{20.77\%} & \text{20.98\%} & \text{21.08\%} & \text{60.34\%} & \text{60.76\%} & \text{60.96\%} \\
\cline{1-9}
\hline
\rowcolor{orange!50} 
\multicolumn{9}{|c|}{\textbf{Shape Schema 3: $|S|=14$, C=112}}\\
\cline{1-9}
\rowcolor{orange!20}
\multicolumn{3}{|c|}{Small Knowledge Graphs (SKGs)} & \multicolumn{3}{|c|}{Medium Knowledge Graphs (MKGs)} & \multicolumn{3}{|c|}{Large Knowledge Graphs (LKGs)}\\
\textbf{D19} & \textbf{D20} & \textbf{D21} & \textbf{D22} & \textbf{D23} & \textbf{D24} &  \textbf{D25} & \textbf{D26} & \textbf{D27}\\
\multicolumn{1}{|c}{\textbf{\#triples:}} & \multicolumn{2}{c|}{1,001,420} & \multicolumn{1}{|c}{\textbf{\#triples:}} & \multicolumn{2}{c|}{4,257,051} &
\multicolumn{1}{|c}{\textbf{\#triples:}} & \multicolumn{2}{c|}{34,095,887}\\
\multicolumn{1}{|c}{\textbf{\#subjects:}} & \multicolumn{2}{c|}{163,553} & \multicolumn{1}{|c}{\textbf{\#subjects:}} & \multicolumn{2}{c|}{693,270} &
\multicolumn{1}{|c}{\textbf{\#subjects:}} & \multicolumn{2}{c|}{5,543,986}\\
\multicolumn{1}{|c}{$\overline{\textbf{inter}}$} & \multicolumn{2}{c|}{7,658.66} & \multicolumn{1}{|c}{$\overline{\textbf{inter}}$} & \multicolumn{2}{c|}{32,611.29} &
\multicolumn{1}{|c}{$\overline{\textbf{inter}}$} & \multicolumn{2}{c|}{261,160.13}\\
\multicolumn{1}{|c}{$\overline{\textbf{intra}}$} & \multicolumn{2}{c|}{14,038.84} & \multicolumn{1}{|c}{$\overline{\textbf{intra}}$} & \multicolumn{2}{c|}{59,663.18} &
\multicolumn{1}{|c}{$\overline{\textbf{intra}}$} & \multicolumn{2}{c|}{477,493.14}\\
\cline{1-6} 
\rowcolor{orange!20}
\multicolumn{3}{|c|}{SKGs Invalid Entities \textbf{\%inv}}& \multicolumn{3}{|c|}{MKGs Invalid Entities \textbf{\%inv}} & \multicolumn{3}{|c|}{LKGs Invalid Entities \textbf{\%inv}}\\
\textbf{D19} & \textbf{D20} & \textbf{D21} & \textbf{D22} & \textbf{D23} & \textbf{D24} &  \textbf{D25} & \textbf{D26} & \textbf{D27}\\
\text{7.61\%} & \text{11.52\%} & \text{15.10\%} & \text{22.42\%} & \text{26.41\%} & \text{30.09\%} & \text{52.48\%} & \text{56.90\%} & \text{61.33\%} \\
\cline{1-9}
\end{tabular}
}
\end{table*}

We empirically study the behavior of Trav-SHACL; it is compared to the state-of-the-art SHACL shapes validator SHACL2SPARQL~\cite{Corman2019b,Corman2019}.
We aim to answer the following research questions:
\begin{inparaenum}[\bf RQ1\upshape)]
    \item What is the effect of validating the shapes following different traversal strategies?
    \item Using a shape network, can the knowledge gained from previously validated shapes be exploited to improve the performance?
    \item What is the impact of the size of the data sources, i.e., do the approaches scale up?
    \item What is the impact of the topology of the constraint network and the selectivity of the shapes?
\end{inparaenum}
Source code and experiment scripts are available at GitHub\footnote{\href{https://github.com/SDM-TIB/Trav-SHACL}{https://github.com/SDM-TIB/Trav-SHACL}}.
In the following, the experimental configuration is described. Finally, the results are analyzed.

\subsection{Experimental Setup}
\textbf{Data Sets and Shapes.}
To the best of our knowledge, there are no benchmarks to evaluate the performance of SHACL validators.
Therefore, we build our test beds based on the accepted and commonly used Lehigh University Benchmark (LUBM)~\cite{Guo2005}.
The LUBM Data Generator\footnote{LUBM Data Generator available at \href{http://swat.cse.lehigh.edu/projects/lubm/}{http://swat.cse.lehigh.edu/projects/lubm/}} is used to create data of different sizes; the small, medium, and large knowledge graphs in \autoref{tab:datasets}.
Based on the classes and properties available in the data, we create one shape per class (shape schema 3 in \autoref{tab:datasets}).
From this full shape schema we also evaluate subsets, referred to as shape schema 1 and shape schema 2.
The originally generated data is modified in such a way that for each of the shape schemas each knowledge graph size has three different percentages of invalid entities.
For the ease of the discussion of the results, we call each combination of shape schema, knowledge graph size, and percentage of invalid entities a \emph{data set}.
This leads to a total of 27 data sets to be evaluated.
For more detailed information about the sizes of the shape schemas and other statistics on the data sets we refer the reader to \autoref{tab:datasets}.

\noindent\textbf{SHACL Engines.}
The baseline of our comparison is the state-of-the-art SHACL shape validator SHACL2SPARQL~\cite{Corman2019b,Corman2019}.
Trav-SHACL is implemented in \emph{Python 3.6} and SHACL2SPARQL in \emph{Java 8}.
Due to the performance differences of the two programming languages, we add a Python implementation of the SHACL2SPARQL approach to our study, named SHACL2SPARQL-py.
We compare nine different configurations of Trav-SHACL. As depicted in \autoref{tab:travshacl:config}, they differ in the heuristics used for the seed shape selection and the traversal strategy used to generate the order in which the shapes of the shape schema are evaluated.
In total, we compare eleven different traversal and validation strategies.
To ensure determinism during the experimental study, the engines add an ORDER BY clause to all queries. This is necessary for the case that an engine does not receive all results of an unselective query due to reaching the maximal answer limit of the SPARQL endpoint.

\begin{table}[!ht]
    \caption{\textbf{Trav-SHACL Configurations.} The traversal strategy is used to find the order in which the shapes are validated. The seed shape is selected from the shapes with the highest in- or outdegree. In the case that two or more shapes belong to this group, the shape with the most or fewest constraints is chosen. Trav-SHACL 9 works completely random.}
    \label{tab:travshacl:config}
    \centering
    \begin{tabular}{cccc}
        \rowcolor{orange!50}
         &&\multicolumn{2}{c}{Seed Shape Selection}\\\rowcolor{orange!50}
         \multirow{-2}{*}{Name} &\multirow{-2}{*}{\shortstack[c]{Traversal\\ Strategy}} &Connectivity &Constraints\\\hline
         Trav-SHACL 1 &BFS &high indegree &many\\
         Trav-SHACL 2 &BFS &high indegree &few\\
         Trav-SHACL 3 &BFS &high outdegree &many\\
         Trav-SHACL 4 &BFS &high outdegree &few\\
         Trav-SHACL 5 &DFS &high indegree &many\\
         Trav-SHACL 6 &DFS &high indegree &few\\
         Trav-SHACL 7 &DFS &high outdegree &many\\
         Trav-SHACL 8 &DFS &high outdegree &few\\
         Trav-SHACL 9 &\multicolumn{3}{c}{random}\\
    \end{tabular}
\end{table}

\noindent\textbf{Metrics.}
We report the following metrics:
\begin{inparaenum}[\bf a\upshape)]
    \item \emph{Average Validation Time}: Average time elapsed between starting the validation of a data set and the engine finishing the validation. It corresponds to absolute wall-clock system time in seconds.
    \item \emph{Standard Deviation}: The standard deviation of the validation time.
    \item {\bf $dief@t$}: A measurement for the continuous efficiency of an engine in the first $t$ time units of the validation~\cite{Acosta2017}. The diefficiency is computed as the area under the curve (AUC) of the answer distribution function. Hence, approaches that produce answers faster in a certain time period show higher diefficiency values.
\end{inparaenum}

\begin{figure}[t]
    \centering
    \includegraphics[width=\linewidth]{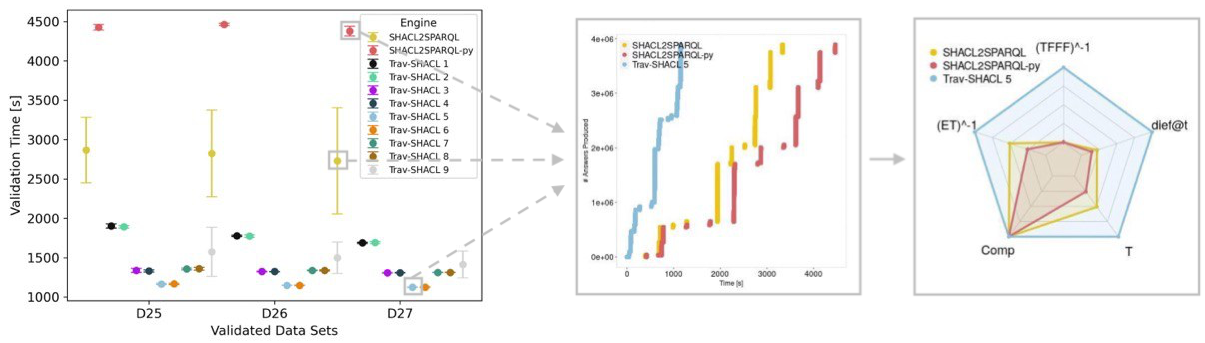}
    \caption{\textbf{Overview of Result Plots.} On the left, validation time in seconds for each data set. Comparison of continuous behavior of best approach with the baseline. In the middle, answer traces showing the incremental generation of validation results. On the right, diefficiency at time $t$. A higher value means a steadier answer production.}
    \label{fig:expresults}
\end{figure}

\noindent\textbf{Experimental setup.}
Each of the 297 experiments is run five times.
An experiment is the validation of a data set D$i$, $i \in [1,27]$ with a particular engine.
All caches are flushed between two consecutive experiments.
All components of the experiment are run in dedicated Docker containers to ensure reproducibility.
We use \emph{Virtuoso 7.20.3229} as SPARQL endpoint for querying the data sets.
All containers are run at the same server.
Hence, network cost can be neglected.
The experiments are executed on an Ubuntu 18.04.4 LTS 64 bit machine with an Intel$^\text{®}$ Xeon$^\text{®}$ E5-1630v4 CPU (four physical cores, eight threads), and 64 GiB DDR4 RAM.
Virtuoso endpoints are configured to use up to 32 GiB while the containers for the SHACL validators are limited to 24 GiB.

\subsection{Discussion of Observed Results}
\begin{figure}[t]
    \centering
    \subfloat[Shape Schema 1 SKGs]{
        \includegraphics[trim=0 0 0 21,clip,width=0.325\linewidth]{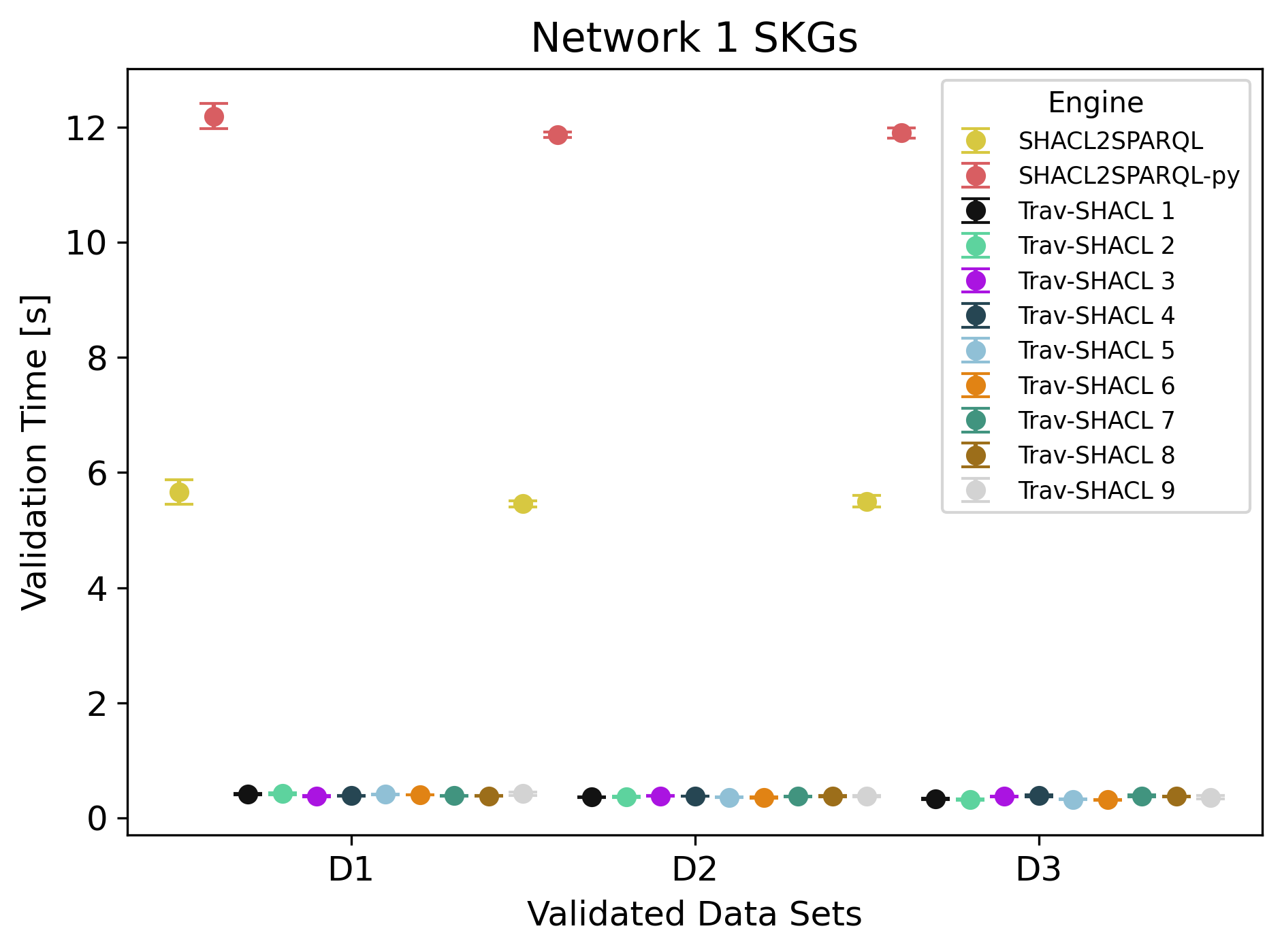}
        \label{fig:time:net1-skg}
    }
    \subfloat[Shape Schema 1 MKGs]{
        \includegraphics[trim=0 0 0 21,clip,width=0.325\linewidth]{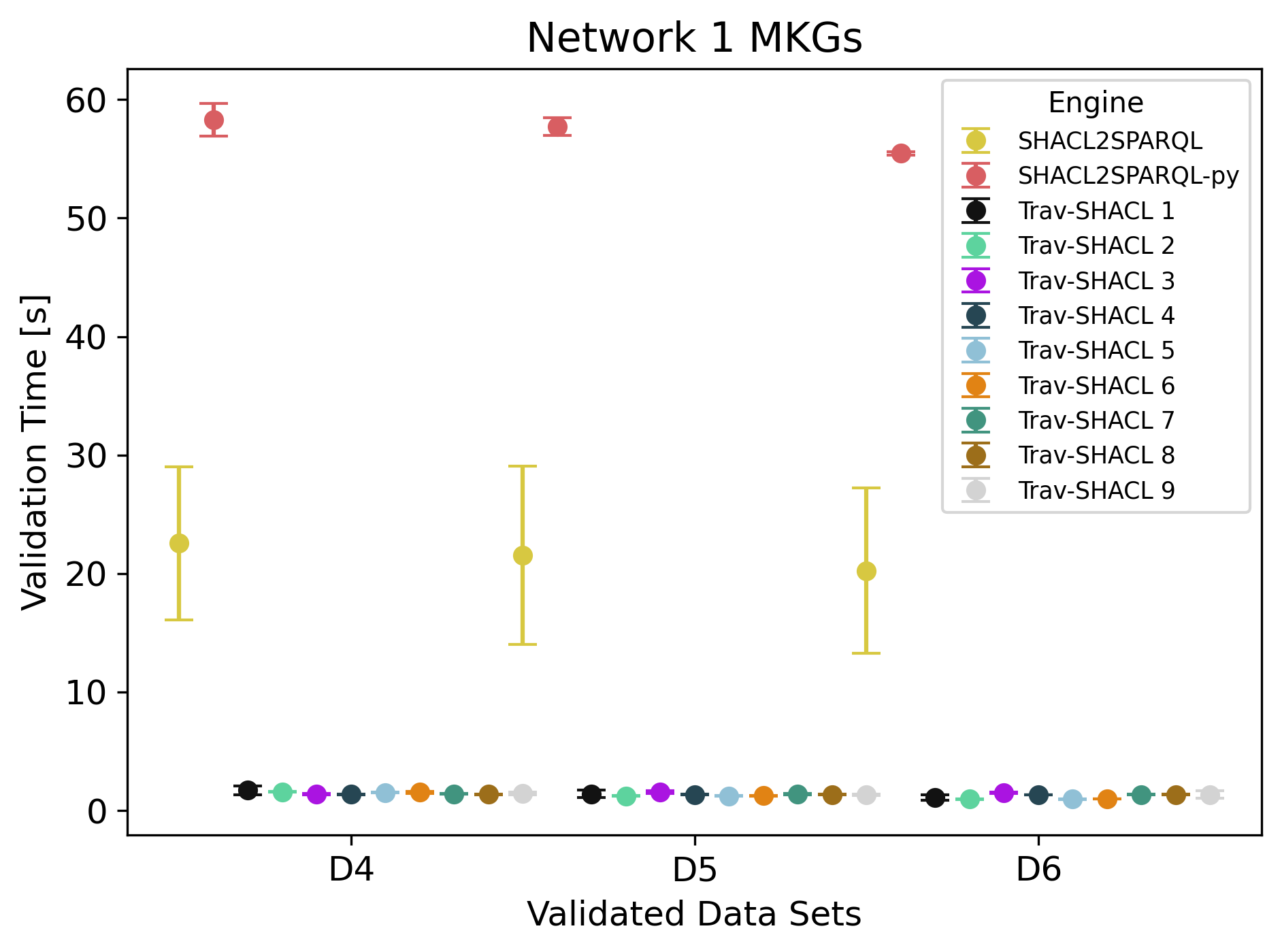}
        \label{fig:time:net1-mkg}
    }
    \subfloat[Shape Schema 1 LKGs]{
        \includegraphics[trim=0 0 0 21,clip,width=0.325\linewidth]{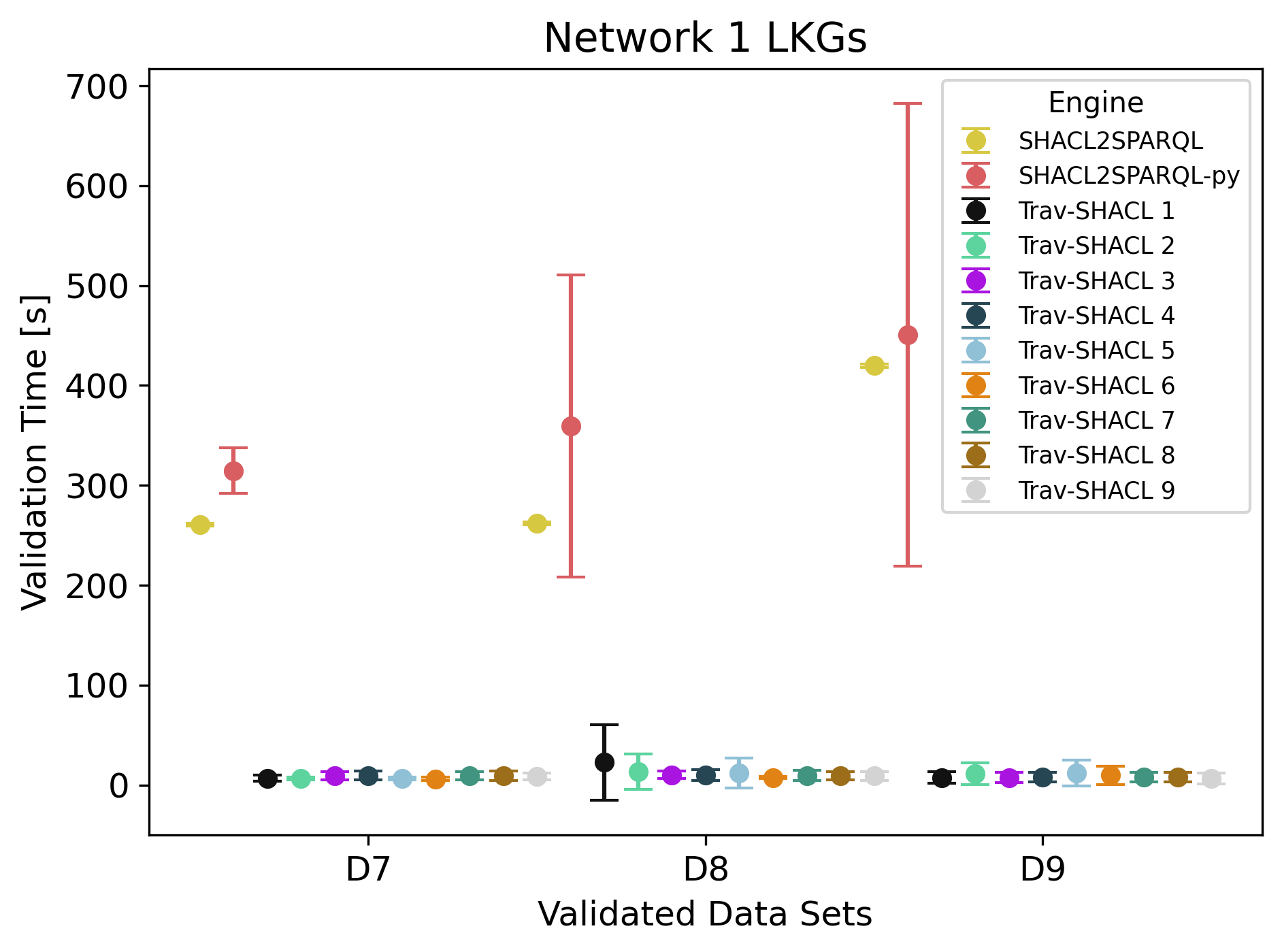}
        \label{fig:time:net1-lkg}
    }
    ~\\\vspace*{.75em}
    \subfloat[Shape Schema 2 SKGs]{
        \includegraphics[trim=0 0 0 21,clip,width=0.325\linewidth]{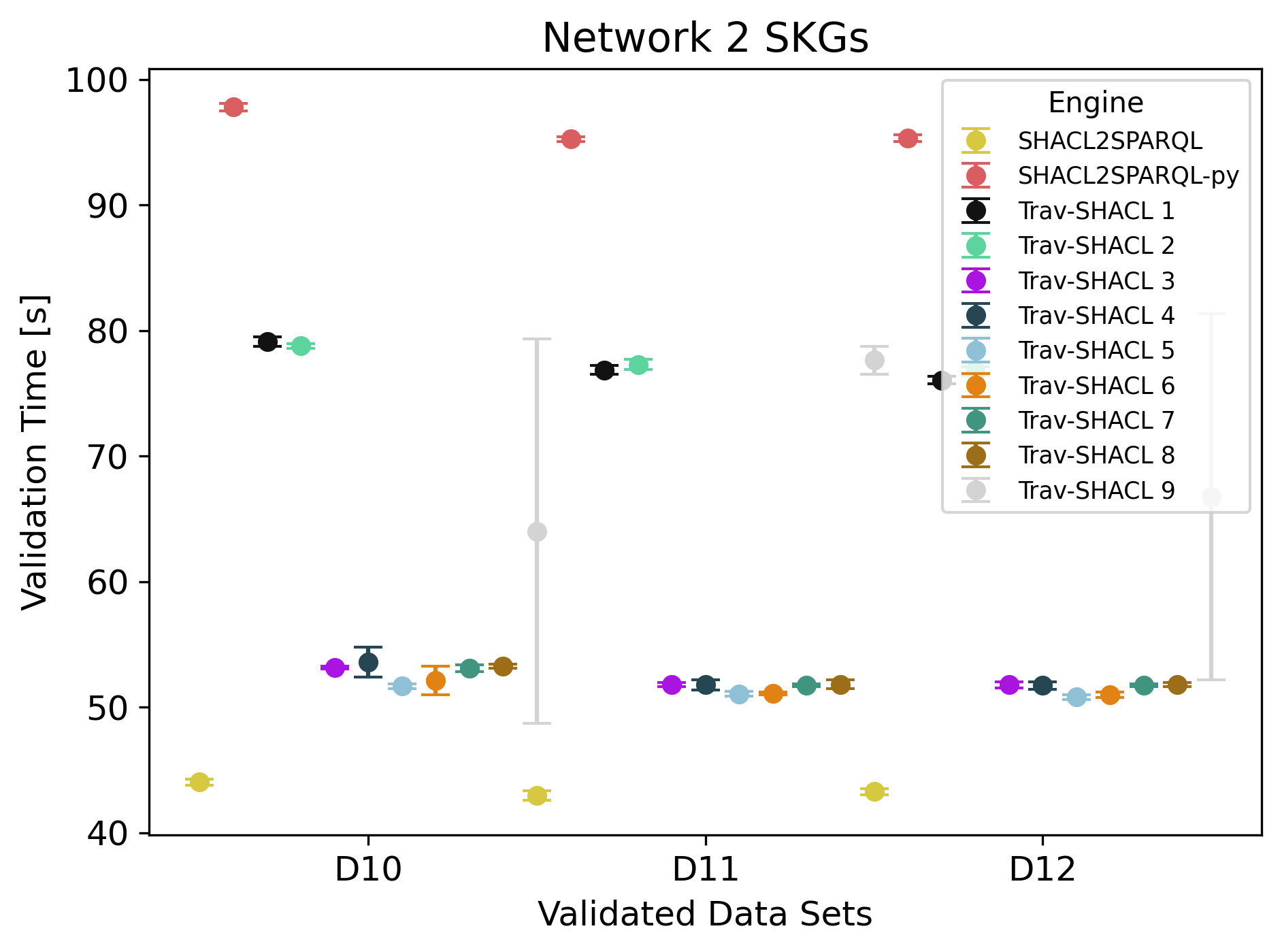}
        \label{fig:time:net2-skg}
    }
    \subfloat[Shape Schema 2 MKGs]{
        \includegraphics[trim=0 0 0 21,clip,width=0.325\linewidth]{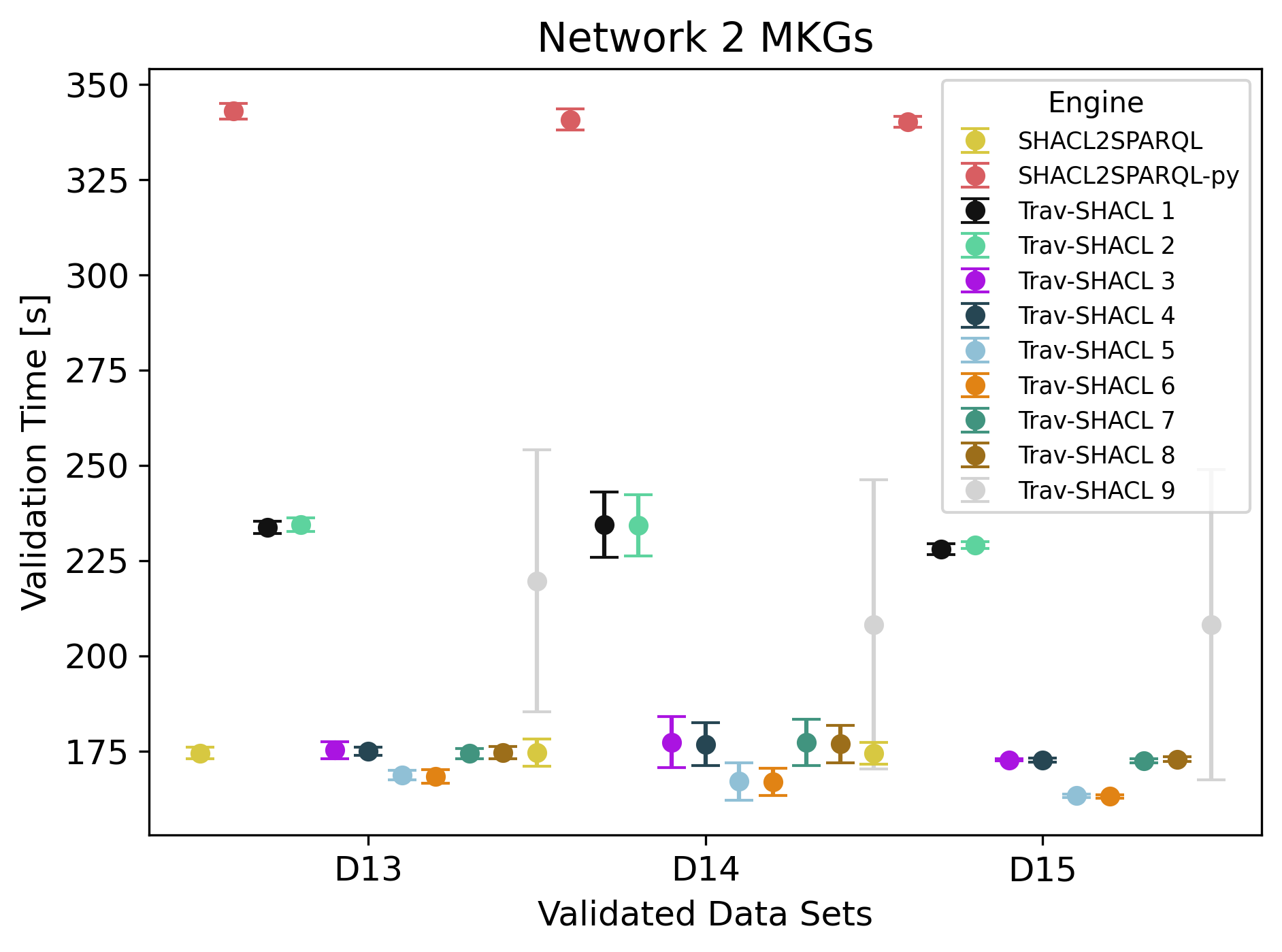}
        \label{fig:time:net2-mkg}
    }
    \subfloat[Shape Schema 2 LKGs]{
        \includegraphics[trim=0 0 0 21,clip,width=0.325\linewidth]{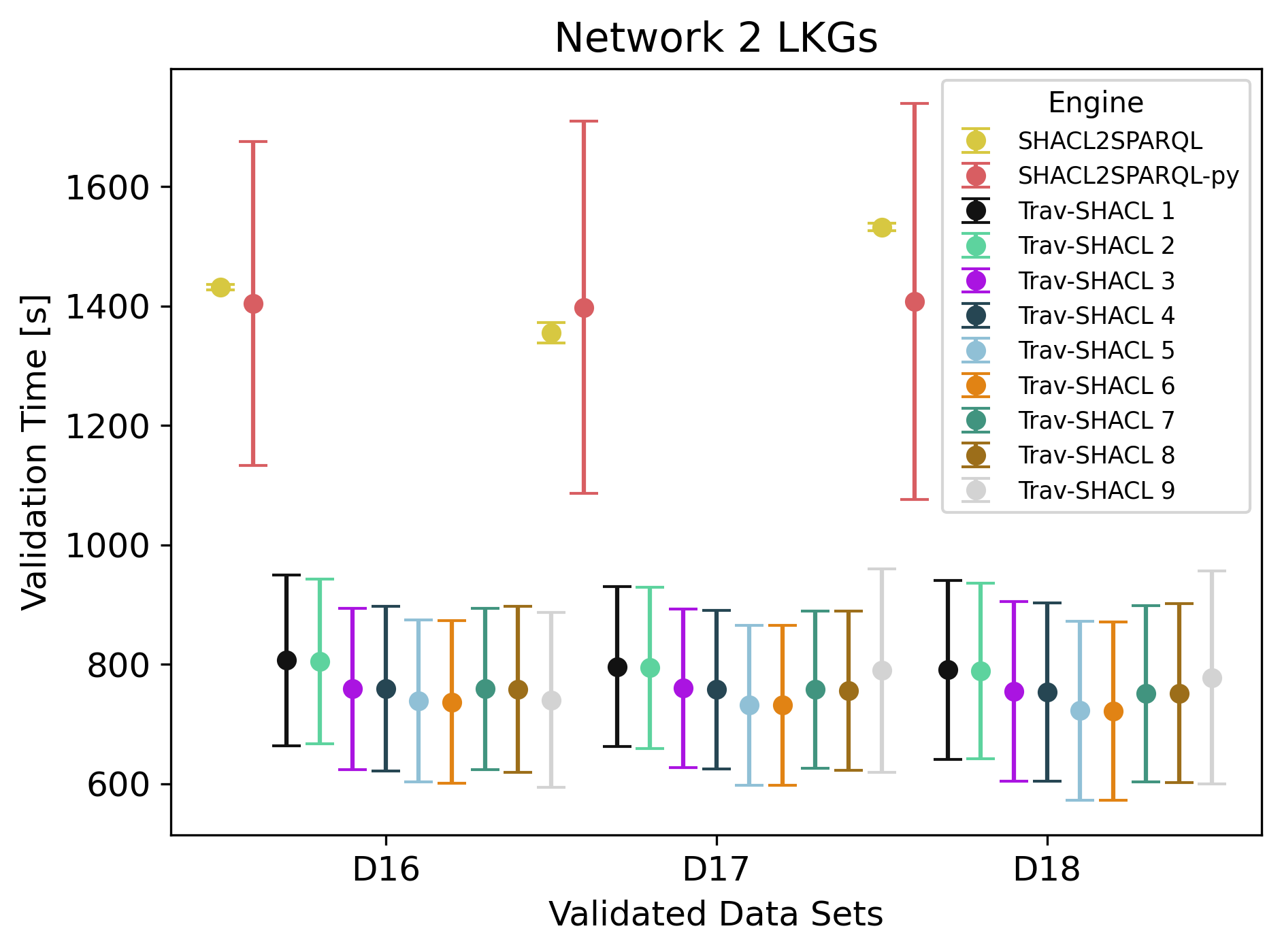}
        \label{fig:time:net2-lkg}
    }
    ~\\\vspace*{.75em}
    \subfloat[Shape Schema 3 SKGs]{
        \includegraphics[trim=0 0 0 21,clip,width=0.325\linewidth]{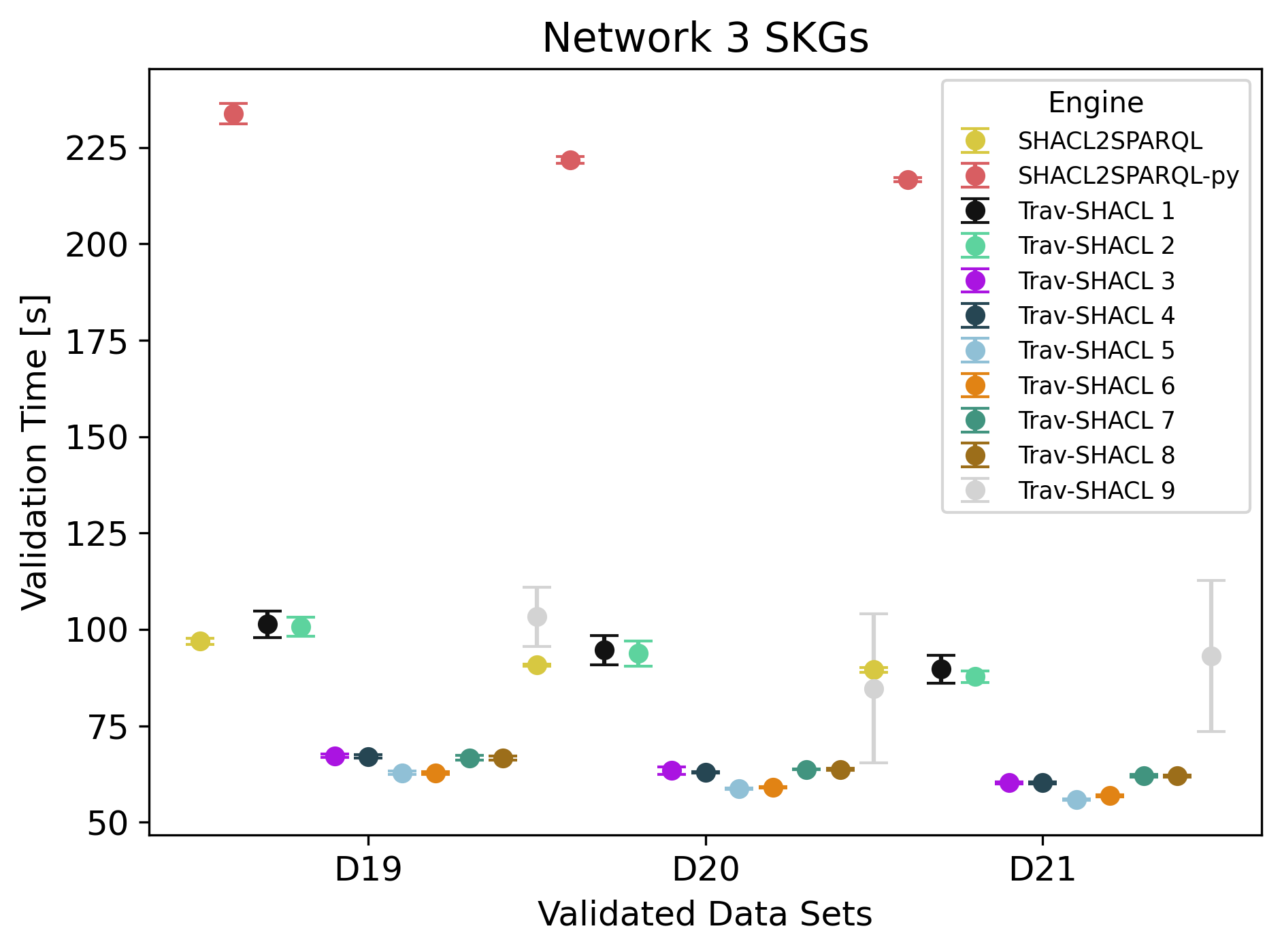}
        \label{fig:time:net3-skg}
    }
    \subfloat[Shape Schema 3 MKGs]{
        \includegraphics[trim=0 0 0 21,clip,width=0.325\linewidth]{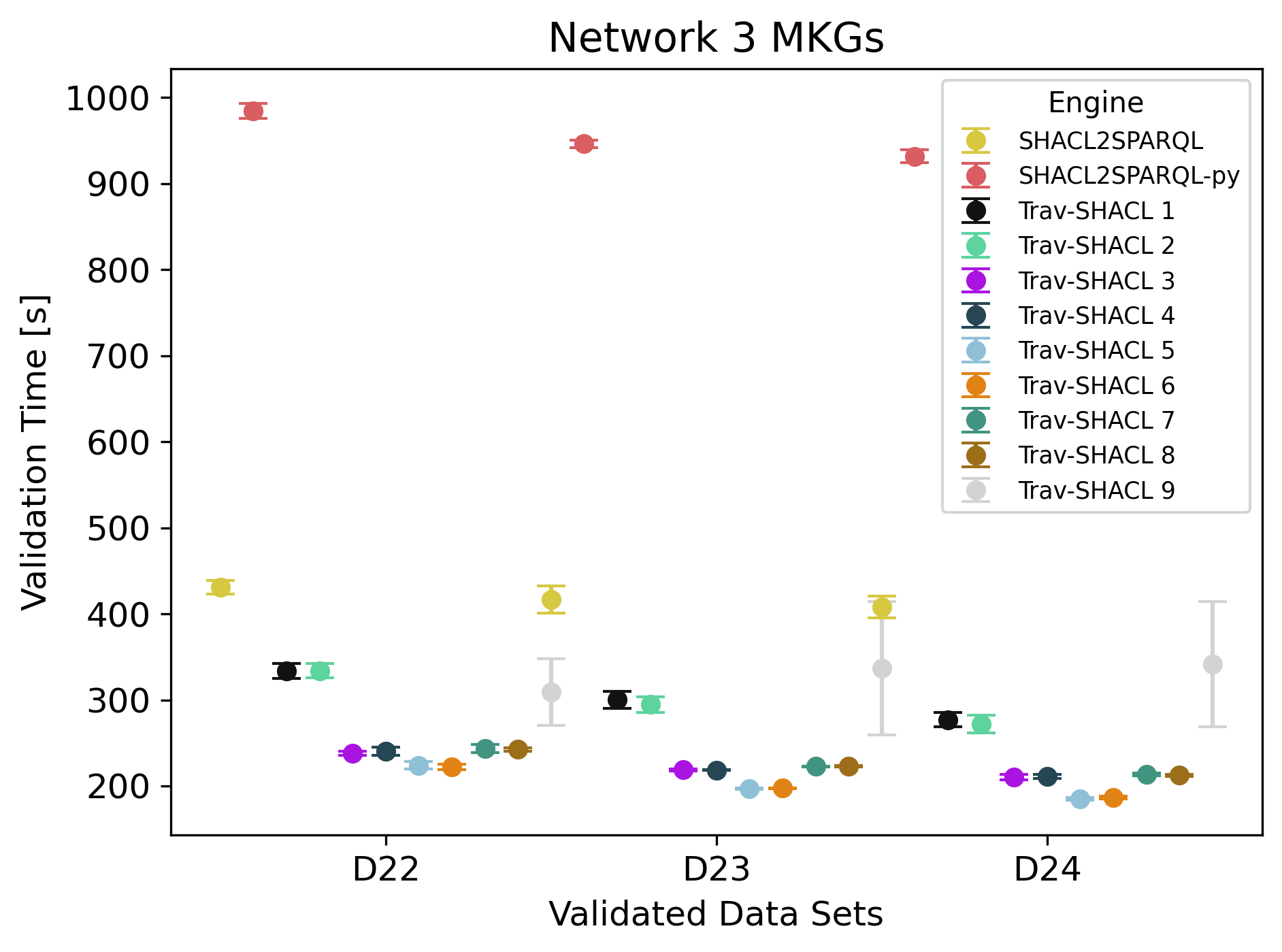}
        \label{fig:time:net3-mkg}
    }
    \subfloat[Shape Schema 3 LKGs]{
        \includegraphics[trim=0 0 0 21,clip,width=0.325\linewidth]{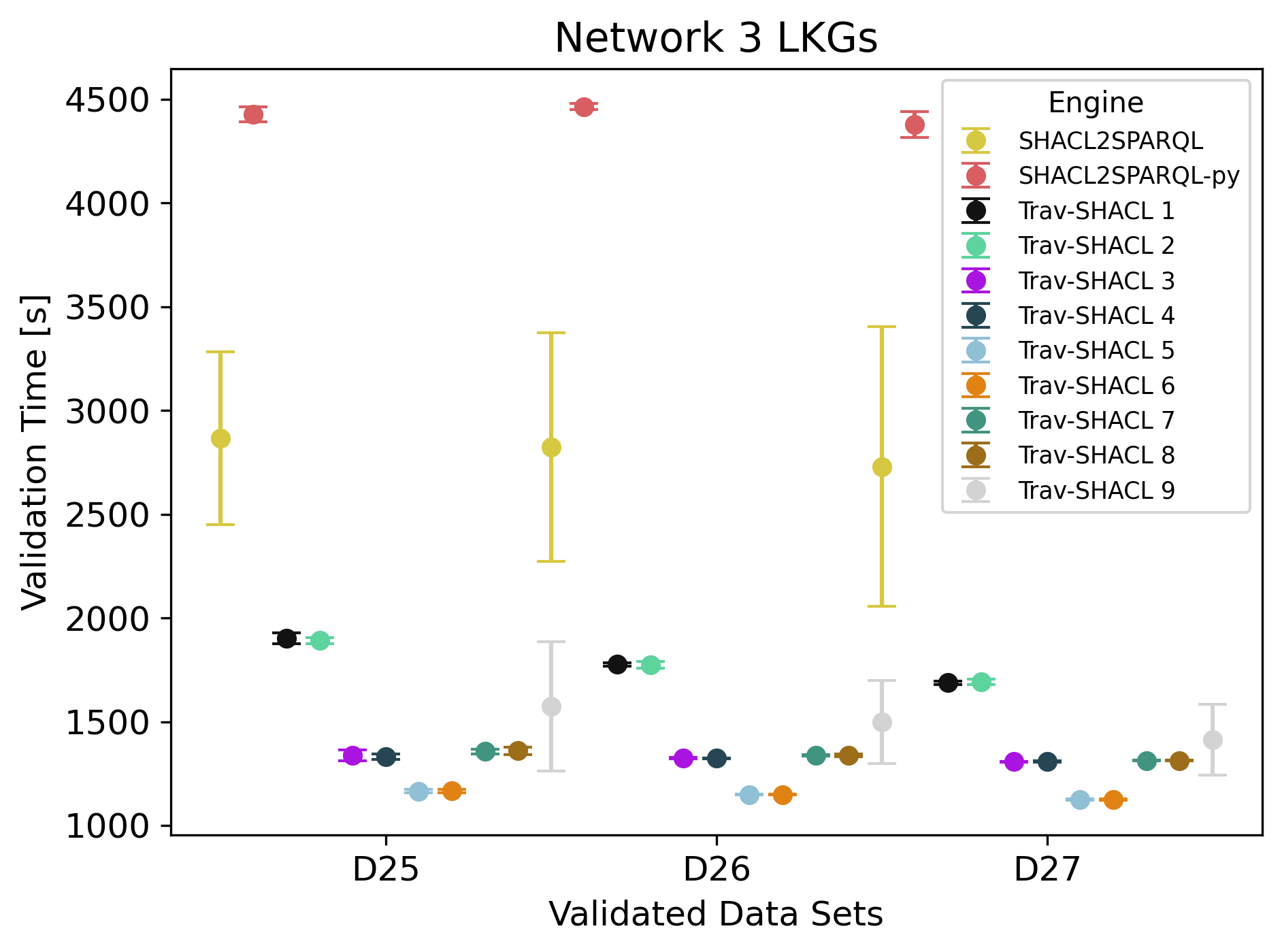}
        \label{fig:time:net3-lkg}
    }
    
    \caption{\textbf{Validation Time of the Experiments in Seconds.} The validation time increases with the size of the data and shape schema. Apart from case (d), Trav-SHACL outperforms SHACL2SPARQL by a factor of up to 28.93.}
    \label{fig:time}
\end{figure}

\begin{figure}[h!]
    \centering
    \subfloat[Shape Schema 1 SKGs]{
        \includegraphics[trim=50 95 35 50,clip,width=.30\linewidth]{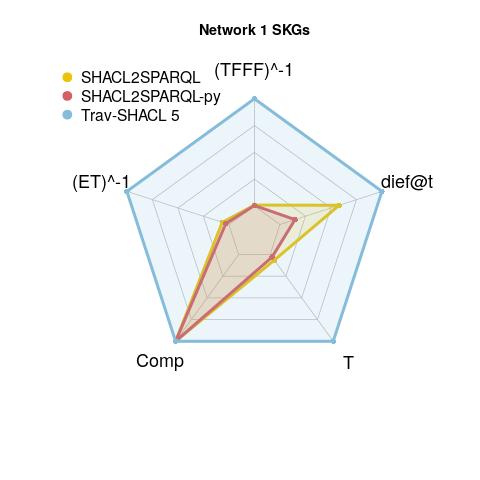}
        \label{fig:dief:net1-skg}
    }\hspace*{.75em}
    \subfloat[Shape Schema 1 MKGs]{
        \includegraphics[trim=50 95 35 50,clip,width=.30\linewidth]{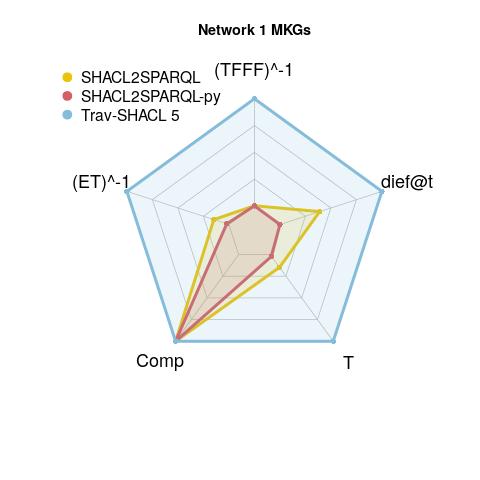}
        \label{fig:dief:net1-mkg}
    }\hspace*{.75em}
    \subfloat[Shape Schema 1 LKGs]{
        \includegraphics[trim=50 95 35 50,clip,width=.30\linewidth]{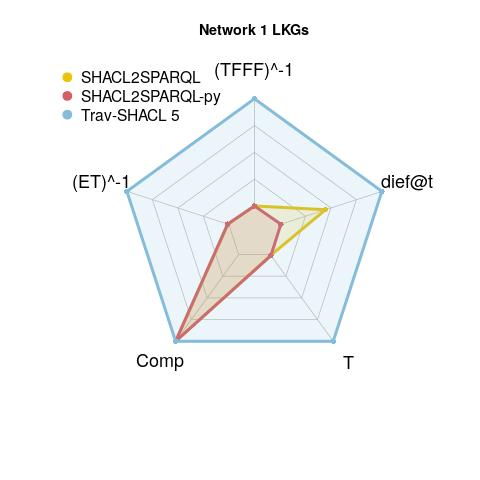}
        \label{fig:dief:net1-lkg}
    }
    ~\\\vspace*{.75em}
    \subfloat[Shape Schema 2 SKGs]{
        \includegraphics[trim=50 95 35 50,clip,width=.30\linewidth]{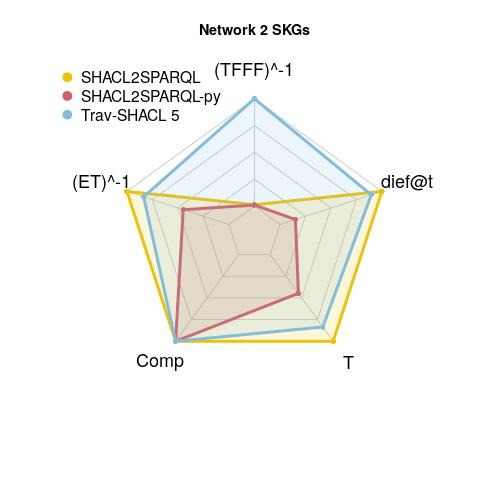}
        \label{fig:dief:net2-skg}
    }\hspace*{.75em}
    \subfloat[Shape Schema 2 MKGs]{
        \includegraphics[trim=50 95 35 50,clip,width=.30\linewidth]{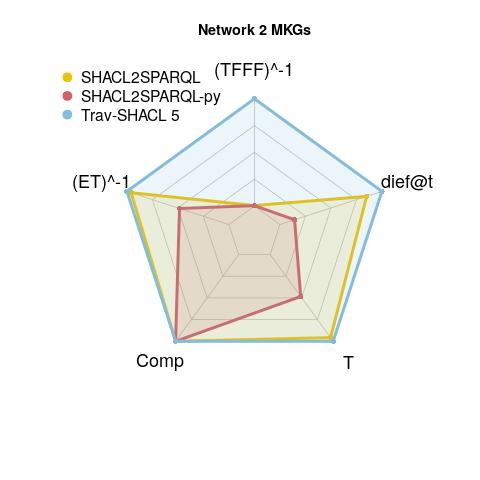}
        \label{fig:dief:net2-mkg}
    }\hspace*{.75em}
    \subfloat[Shape Schema 2 LKGs]{
        \includegraphics[trim=50 95 35 50,clip,width=.30\linewidth]{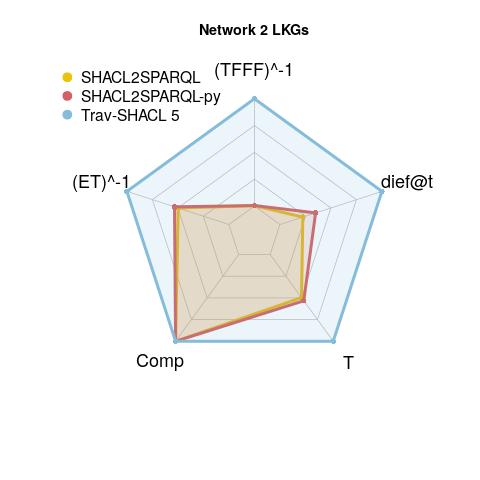}
        \label{fig:dief:net2-lkg}
    }
    ~\\\vspace*{.75em}
    \subfloat[Shape Schema 3 SKGs]{
        \includegraphics[trim=50 95 35 50,clip,width=.30\linewidth]{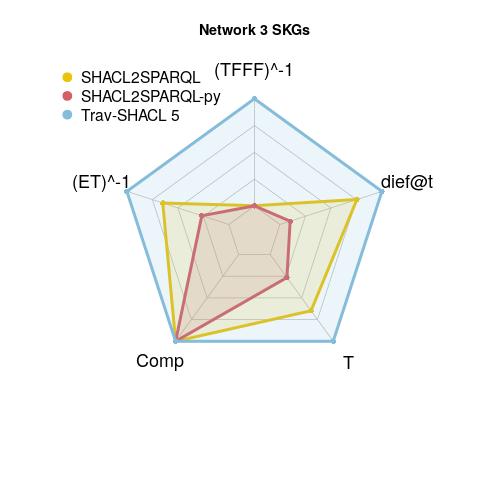}
        \label{fig:dief:net3-skg}
    }\hspace*{.75em}
    \subfloat[Shape Schema 3 MKGs]{
        \includegraphics[trim=50 95 35 50,clip,width=.30\linewidth]{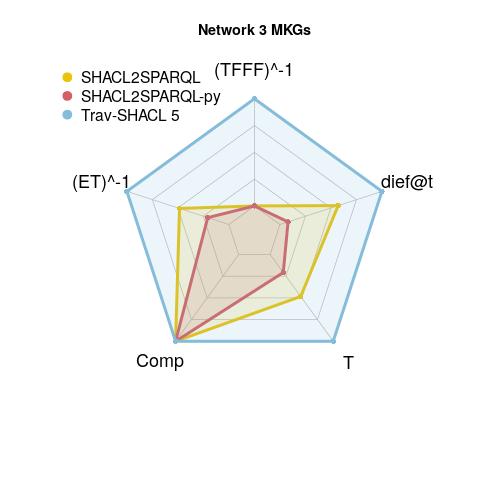}
        \label{fig:dief:net3-mkg}
    }\hspace*{.75em}
    \subfloat[Shape Schema 3 LKGs]{
        \includegraphics[trim=50 95 35 50,clip,width=.30\linewidth]{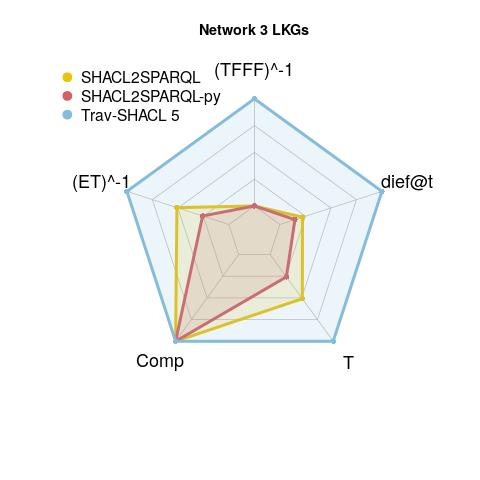}
        \label{fig:dief:net3-lkg}
    }
    \caption{\textbf{Continuous Behavior for the Experiments.} In all cases, Trav-SHACL produces the first result ahead of the other engines. In most cases, Trav-SHACL finishes the validation the fastest. With one exception in (d), Trav-SHACL outperforms the other engines in terms of continuously delivering results. The difference increases with larger knowledge graphs. $(TFFF)^{-1}$ - inverse time for the first answer produced, $(ET)^{-1}$ - inverse validation time (a.k.a. execution time), Comp - sum of validated and invalidated entities, T - throughput, and $dief@t$ - continuous efficiency at time $t$}
    \label{fig:dief}
\end{figure}

\autoref{fig:expresults} gives an overview of the plots presented for the result analysis.
First, we report on the average validation time including a visualization of the standard deviation. Each of these plots shows the validation time of all eleven engines, validating three data sets that only differ in the percentage of invalid entities as depicted in \autoref{tab:datasets}, i.e., the data sets are of the same size.
During the experiments, we collect the timestamp for each validation decision made.
From those timestamps, we generate the answer trace. The traces show the continuous generation of answers for the engines. For the sake of readability, we only include the overall best configuration of Trav-SHACL, namely Trav-SHACL 5.
Finally, we present the diefficiency at time $t$ ($dief@t$). All measurements presented in the radar plot are 'higher the better'. On top, the inverse time for the first produced answer is noted. Hence, a higher values means the first answer was produced faster. To the left, the inverse validation time is reported. Again, a higher value implies earlier termination of the task. In the left bottom corner, Comp refers to the number of validations performed, i.e., the sum of all validated and invalidated entities. Next, the throughput T indicates the speed with which the answers are produced. A higher value implies shorter intervals between the report of two validation results. Finally, the diefficiency at time $t$, if engine $e_1$ has a higher value for $dief@t$ than $e_2$, $e_1$ exhibits a better performance until $t$ in terms of diefficiency.

\noindent\textbf{Validation Time.}
The validation time of the eleven engines for each of the 27 data sets is depicted in \autoref{fig:time}.
\autoref{fig:time:net1-skg} reports on the validation time for the small knowledge graphs against shape schema 1. As can be seen, the Python version of SHACL2SPARQL performs worst. Trav-SHACL outperforms SHACL2SPARQL-py by a factor of 31.58 and SHACL2SPARQL by a factor of 14.47. There is no considerable difference in the configurations of Trav-SHACL.
The time needed to validate shape schema 1 against the medium-sized knowledge graphs is presented in \autoref{fig:time:net1-mkg}. We observe a very similar behavior compared to the smaller knowledge graphs.
For the large knowledge graphs, shown in \autoref{fig:time:net1-lkg}, we start to see a small difference in the times reported for the configurations of Trav-SHACL. Also the factors by which we outperform the other approaches change. Trav-SHACL outperforms SHACL2SPARQL by a factor of 28.93 and SHACL2SPARQL-py by 40.06.
The validation of shape schema 2 against small knowledge graphs shows a different behavior as can be seen in \autoref{fig:time:net2-skg}. This is because of the small amount of invalid entities in the data sets (see \autoref{tab:datasets}). This leads to the fact that SHACL2SPARQL outperforms Trav-SHACL by a factor of 1.18 because Trav-SHACL cannot take advantage by its capability of discovering invalid entities fast.
For the medium-sized knowledge graphs, Trav-SHACL is able to exploit its capabilities and perform similarly. Trav-SHACL 5 is the best approach in that case. However, it outperforms SHACL2SPARQL only by a factor of 1.04.
As in the previous shape schema, the difference in validation time becomes more clear in the presence of large knowledge graphs. Trav-SHACL outperforms SHACL2SPARQL by a factor of up to 2.33.
The high variance in validation for Trav-SHACL in \autoref{fig:time:net2-lkg} is caused by one single outlier in the five runs for each of the configurations. It is always the same query for the big shape \emph{UndergraduateStudent} that takes about twice as long as in the other runs. We observe this behavior at the SPARQL endpoint even though all caches are flushed between experiments.
Moving to the validation of shape schema 3, Trav-SHACL outperforms the other engines.
In the case of small knowledge graphs, Trav-SHACL is up to 1.57 times faster than SHACL2SPARQL.
When validating the shape schema against a medium-sized knowledge graph, we perform up to 2.20 times faster.
Considering large knowledge graphs, Trav-SHACL outperforms SHACL2SPARQL by a factor of up to 2.46.
To sum up, Trav-SHACL is outperforming SHACL2SPARQL. It can be observed that the factor by which we perform better is increasing with larger knowledge graphs. However, the size of the knowledge graph is not the only impacting factor. The properties of the shape network, e.g., the cardinality of the shapes and percentage of invalid entities play an important role. The best engine overall is Trav-SHACL 5.

\noindent\textbf{Continuous Behavior.}
We report the continuous behavior of the eleven engines for each group of data sets instead of the single data sets since the behavior within a group does not change.
As discussed above, the results are visualized using radar plots in \autoref{fig:dief}.
\autoref{fig:dief:net1-skg} to \autoref{fig:dief:net1-lkg} present the results for shape schema 1.
Trav-SHACL outperforms the other engines in all measurements; all the engines identify the same number of (in)valid entities (Comp) except in the LKGs. Even though, the shape schema is not very complex, SHACL2SPARQL and SHACL2SPARQL-py perform similar but in the diefficiency at time $t$. This observation leads to the conclusion that the Java implementation is able to produce the validated answers at a faster rate.
Moving to shape schema 2, \autoref{fig:dief:net2-skg} shows that Trav-SHACL is able to produce the first result faster than SHACL2SPARQL. However, due to the low amount of invalid entities, Trav-SHACL is not able to exploit its advantages, leading to a worse validation time and diefficiency at $t$. When validating the shape schema against medium-sized knowledge graphs, the execution times are competitive. But Trav-SHACL produces the answers steadier than SHACL2SPARQL. In the case of large knowledge graphs, Trav-SHACL outperforms SHACL2SPARQL in all reported measurements.
Considering shape schema 3, Trav-SHACL outperforms all engines in all measurements for all knowledge graph sizes (see \autoref{fig:dief:net1-skg} - \autoref{fig:dief:net1-lkg}). The difference between Trav-SHACL and SHACL2SPARQL increases with larger knowledge graphs. The results show that the Java implementation of SHACL2SPARQL outperforms the Python implementation of the same approach.
To sum up, Trav-SHACL exhibits a better performance in terms of the diefficiency at time $t$ than SHACL2SPARQL.

\noindent\textbf{Correctness of Validation.}
All engines produce the same results for all shape schemas validated against SKGs and MKGs.
However, SHACL2SPARQL fails to correctly validate the shape schemas over LKGs.
In order to enable the comparison of the (in)validated entities between the three prototypical implementations studied in this paper, we ordered constraint queries by the common subject of all query triples, with the usage of the \textsc{ORDER BY} clause. This is necessary since the number of answers retrieved by the SPARQL endpoint for a query is bounded.
Then, when evaluating large knowledge graphs like D25 (see \autoref{tab:datasets}),
SHACL2SPARQL is not able to identify 41.10\% of the valid entities with respect to Trav-SHACL, but classifies them as invalid instead. This is due to the lack of selectivity in the definition of the constraint queries. Contrary, Trav-SHACL faces the same external limitations imposed by the SPARQL endpoint, but it leverages the query rewriting strategy to retrieve more relevant data for the validation process.

\noindent\textbf{Answer to RQ1.}
From the analysis of the results, it is clear that the traversal strategy impacts on the validation time. Naturally, the difference is more prominent in shape schemas with more shapes. But also the size of the data matters as the difference increases also with larger knowledge graphs. The analysis shows that overall Trav-SHACL 5 performs best. The seed shape is selected as the one with the most constraints amongst the shapes with the highest indegree. That allows for a reuse of the validation results of the first shape for many of the following shapes. The results of our study prove the intuition mentioned when describing the \emph{Inter-shape planner}.

\noindent\textbf{Answer to RQ2.}
The knowledge gained from previously validated shapes can be exploited to improve the performance of the validation engine. This can be seen in the great savings of Trav-SHACL compared to SHACL2SPARQL-py and in the decreasing validation time of Trav-SHACL with increasing number of invalid entities.

\noindent\textbf{Answer to RQ3.}
Trav-SHACL scales up, and its performance difference to SHACL2SPARQL increases with larger knowledge graphs. Trav-SHACL is able to produce correct results but SHACL2SPARQL failed to correctly classify the entities in the large knowledge graphs. This is caused by a non-selective constraint query; therefore, SHACL2SPARQL does not receive all the answers needed for the validation as discussed in \textbf{Correctness of Validation}.

\noindent\textbf{Answer to RQ4.}
The validation time increases with an increasing number of shapes in a shape schema.
However, more important than the number of shapes is the cardinality of the shapes, i.e., the number of entities in $[[\textsc{targ}(s)]]^\mathcal{G}$. For example, the difference in validation time of shape schema 2 and shape schema 3 against the same knowledge graph is relatively small taking into consideration that shape schema 3 consists of twice as many shapes. This is due to the fact that the two biggest shapes in terms of entities assigned to them, are already included in shape schema 2. Those two shapes comprise almost 70\% of all entities in the data. Therefore, validating many selective shapes is faster than the validation of a single shape with many entities assigned to it. Trav-SHACL also benefits in the cases where many shapes refer to the same shape. To conclude, the ideal case is many selective shapes that all depend on one other shape or, alternatively, a chain of shapes.

\section{Related Work}\label{sec:relatedwork}
\textbf{Constraint Languages.} Several approaches have been developed towards the definition and evaluation of expressive constraints on semantic web models. Initial works correspond to the definition of integrity constraint semantics in OWL by using the closed-world assumption~\cite{Motik2007,Motik2009,Tao2010}. However, OWL was originally designed to model incomplete data with the open-world assumption, hence, not well-suited for expressing integrity constraints.
A next step was the SPARQL Inferencing Notation (SPIN)~\cite{SPIN}, a W3C member submission that suggested the use of SPARQL queries as constraints on top of RDF graphs.
The next generation of SPIN is the W3C recommendation language SHACL~\cite{SHACL}, a language that allows to represent integrity constraints over RDF graphs in RDF.
ShEx~\cite{Thornton2019} is a constraint language for RDF and inspired by schema languages for XML. ShEx is similar to SHACL but the semantics of ShEx builds on regular bag expressions. While SHACL semantics allows multiple possible assignments, a single assignment is created when validating an RDF graph with ShEx. It is worth mentioning that any graph that is valid with respect to ShEx shapes will be valid to an equivalent constraint definition in SHACL while this does not hold in the other direction.
Validation algorithms with semantics for recursion with stratified negation have been proposed by Boneva et al. \cite{Boneva2017} for ShEx.
Since SHACL is the W3C recommendation, we focus on the validation of integrity constraints expressed in SHACL.
~\\
\noindent\textbf{Shape Modeling.}
The above-mentioned languages for representing integrity constraints over RDF graphs are a foundation for quality assessment. As a first step in a quality assessment pipeline, a shape schema for the data to validate needs to be found. Many works have been done in automatic shape generation from the data or systems that guide the data administrator in the process of generating the constraints in a semi-automatic manner.
ABSTAT~\cite{Spahiu2016} is an online semantic profiling tool for data-driven extraction of ontology patterns and data statistics. Data consumers can benefit from ABSTAT in better understanding the data. Spahiu et al.~\cite{Spahiu2018} propose a methodology to transform the ABSTAT profiles into SHACL for quality assessment. The SHACL constraints are based on the patterns discovered by ABSTAT. Therefore, most of the constraints are cardinality constraints or domain/range constraints.
In contrast to ABSTAT, Astrea~\cite{Cimmino2020} generates SHACL shapes from ontologies only. Astrea uses mappings between ontology patterns and SHACL patterns for automatic shape generation. The ontology patterns are extracted from OWL 2, RDFS, and XSD.
The modeling of a shape schema for a given RDF graph is beyond the scope of this paper.
~\\
\noindent\textbf{SHACL Validation.}
Another important step in a quality assessment pipeline is the actual execution of the validation of the shape schema against the knowledge graph.
The validation of recursive shape schemas is left undefined in the specification of SHACL~\cite{SHACL}. Corman et al.~\cite{Corman2018} introduce a semantics for the validation of resursive SHACL. They also show that the validation of the full SHACL features is NP-hard. Based on those findings, they propose fragments of SHACL that are tractable together with a basic algorithm for validating a shape schema using SPARQL~\cite{Corman2019}. Andreşel et al.~\cite{Andresel2020} introduce an either stricter semantics for recursive SHACL based on stable models known from Answer Set Programming (ASP). This approach allows to represent SHACL constraints as logic programs and use existing ASP solvers for the validation of the shape schema. Another advantage is that negations in recursions are possible following the proposed approach. In contrast to these logic approaches to the problem, we use query optimization techniques to improve the incremental behavior and scalability.
~\\
\noindent\textbf{Satisfiability and Containment.}
Recent work focuses on the satisfiability and containment of SHACL. Leinberger et al.~\cite{Leinberger2020} propose to use description logics for the containment of a shape in another. They study standard entailment for different fragments of SHACL.
Pareti et al.~\cite{Pareti2020} propose a new fragment of first order logic (FOL) extended with counting quantifiers and transitive closure operator called \textsc{SCL} to decide the satisfiability of a shape schema and the containment of a shape schema in another. This fragment only covers the \emph{core constraint components} from the SHACL specification and no recursion. Which makes both problems decidable which they are not for full SHACL. While the problem of shape schema containment is important for data integration, the satisfiability problem is of more interest for quality assessment.

\section{Conclusions and Future work}\label{sec:conclusion}
We addressed the problem of minimizing the execution time of data quality assessment constraints expressed in a SHACL shape schema. Given the increasing acceptance of knowledge graphs at industrial and scientific organizations and the W3C recommendation of SHACL as the language to define integrity constraints against RDF graphs, scalable SHACL engines are necessary for global adoption. 
We presented Trav-SHACL as an effective data management tool to fulfill the scalability requirements of novel knowledge-driven developments. Trav-SHACL selects the traversal shape plans and rewrites the target and constraint queries to the fast detection of invalid entities. In doing so, Trav-SHACL is able to reduce the number of constraints that need to be checked during the validation and produce results incrementally. As a result, total execution time is reduced by a factor of up to 28.93, and results are delivered continuously. 
Thus, Trav-SHACL broadens the repertoire of tools for declaratively creating and curating knowledge graphs. We hope that our reported results encourage the diverse communities to develop applications where these results can be reproduced and generalized in real-world scenarios. 
Since Trav-SHACL aims at invalidating entities fast, it does not perform as well in data sets with a small percentage of invalidated entities as in low-quality data sets. In the future, we plan to investigate adaptive planning techniques able to adjust shape validation schedules to the characteristics of both the shape schema and the knowledge graph. Lastly, the incorporation of Trav-SHACL in real-world pipelines of knowledge graph management is part of our future agenda. 

\section*{Acknowledgements}
This work has been partially supported by the EU H2020 projects iASiS (No 727658) and QualiChain (No 822404), and the ERAMed project P4-LUCAT (No 53000015).

\printbibliography
\end{document}